\documentclass[preprint,prb,aps,showkeys,amsmath,amssymb]{revtex4-1}
\usepackage{epsfig}
\usepackage{amsmath}

\usepackage{epsfig}
\usepackage{amsmath}
\usepackage{color}

\usepackage{color}
\usepackage[latin9]{inputenc}
\usepackage{float}
\usepackage{natbib}
\usepackage{graphicx}

\begin{document}

\title{A force field of Li$^{+}$, Na$^{+}$, K$^{+}$, Mg$^{2+}$, Ca$^{2+}$, Cl$^{-}$, and SO$^{2-}_{4}$ in aqueous solution based on the TIP4P/2005 water model and scaled charges for the ions}

\author{I. M. Zeron}
\author{J. L. F. Abascal}
\author{C. Vega}
\email{cvega@quim.ucm.es} 
\affiliation{Depto. Qu\'{\i}mica F\'{\i}sica I, Fac. Ciencias Qu\'{\i}micas,
Universidad Complutense de Madrid, 28040 Madrid, Spain}

%
\begin{abstract}
In this work, a force field for several ions in water is proposed.  In
particular we consider the cations Li$^{+}$, Na$^{+}$, K$^{+}$, Mg$^{2+}$,
Ca$^{2+}$ and the anions Cl$^{-}$, and SO$_{4}^{2-}$. These ions were selected
as they appear in the composition of seawater and they are also found in
biological systems. The force field proposed (denoted as Madrid-2019) is
non-polarizable and both water molecules and sulfate anions are rigid. For
water we use the TIP4P/2005 model. The main idea behind this work is to further
explore the possibility of using scaled charges for describing ionic solutions.
Monovalent and divalent ions are modeled using charges of 0.85 and 1.7,
respectively (in electron units). The model allows a very accurate description
of the densities of the solutions up to high concentrations.  It also gives
good predictions of viscosities up to 3 molal concentrations.  Calculated
structural properties are also in reasonable agreement with experiment. We have
checked that no crystallization occurred in the simulations at concentrations
similar to the solubility limit. A test for ternary mixtures shows that the
force field provides excellent performance at an affordable computer cost. In
summary, the use of scaled charges, which could be regarded as an effective and
simple way of accounting for polarization (at least to a certain extend),
improves the overall description of ionic systems in water. However for 
purely ionic systems scaled charges will not 
describe adequately neither the solid nor the melt.  
\end{abstract}

\maketitle

\section{Introduction}
On Earth, the overwhelming majority of water is found in oceans and seas.
For this reason, more often than not, samples of water always contain 
a certain amount of solved salts. Since life started in the oceans 
it is not surprising that many of the ions present in the seas are also
found in the living cells, though at lower concentrations.
For this reason, it is clear that understanding ionic solutions is of fundamental 
interest in, at least, two important systems: oceans and living organisms.
For these systems the typical conditions of interest are not too far away from
the room temperature and room pressure (or for moderate pressures up to 1400
bar in deep water). However the composition often changes from one system to
another.  For this reason it is not possible to have experimental measurements
of ionic solutions for a wide range of pressures, temperatures and
compositions.  Computer simulations could be useful to predict some of the
properties and to understand the outcome of the experimental measurements.

In computer simulations the key factor controlling the accuracy of the
predictions is the form of the intermolecular potential between the different
species. This is usually denoted as the force field. In the 70's and 80's it
was common to model ionic solutions using a implicit description of water. In
this description, water was not included in the simulations, but rather its
presence was taken into an implicit way by scaling the coulombic interactions
between ions by the dielectric constant of water. Certainly in this treatment,
simulations were quite fast, since water is the most abundant component in 
ionic solutions.
This approach may be justified at very low concentrations when the ions are
quite far apart but it will not work for moderate or high concentrations when
the ions are closer. With the increase in computer power the simulation of
ionic systems using atomistic models of water became more popular in the last
30 years.

Until recently, force fields for ionic solutions in water were mainly designed to
reproduce experimental values of the hydration free energies and the first
peak of the radial distribution functions (RDFs). In some cases, the lattice
energies of the ionic crystal or quantum-mechanical data on ion-water clusters in
the gas-phase were also taken into account. Polarizable and non-polarizable
force fields have been proposed in the literature for alkali and alkaline earth
halides.%
\cite{smith18,opls,str:jcp88,aqvist,dang:jcp92,beg:jcp94,SmithDang,roux:bj96,peng:jcpca97,wee:jcp03,Jorgensen,lam:jcpb06, ale:pre07,len:jcp07,joung08,gallo_anomalies,cal:jpca10,yu:jctc10,reif:jcp11,gee:jctc11, deu:jcp12, mao:jcp12,mam:jcp13,mou:jctc13,kiss:jcp14,kol:jcp16,elf:epjst16,pethes17} 
Quite often the force fields are based on non-polarizable water models
as SPC,\cite{spc} SPC/E,\cite{spce} TIP3P,\cite{jorgensen83}
TIP4P,\cite{jorgensen83} TIP4P-Ew,\cite{tip4p-ew} TIP4P/2005\cite{abascal05b}
but there are also polarizable models as POL1,\cite{POL1} RPOL,\cite{dang-RPOL}
SWM4-DP,\cite{lamoureuxSWM4DP,lam:cpl06} B3K.\cite{kissB3k}
An overview of the force fields available up to 2011 can be found in Table I of
the paper by Reif and Hunenberger\cite{reif:jcp11} and a set of possible target
properties to be used for the validation of potential parameters is shown
in Fig.~5 of the cited reference. A more recent review by Nezbeda et 
al.,\cite{nezbeda16} focussed on aqueous sodium chloride, discuss in detail
the importance of the training set properties used to fit the force field
parameters.

The need of a more broader check of the properties of these model systems was
clearly underlined by Joung and Cheatham.\cite{joung08} These authors
reported the formation of salt crystals well below the experimental saturation limit in
simulations of alkali halide aqueous solutions using different force fields.
In fact, precipitation in NaCl,\cite{maz:jacs03, ale:pre07, ale:jcp09, mou:jpcb12, mou:jcp13}
KCl,\cite{joung08,auf:jctc07} CaCl$_{2}$,\cite{mar:jcp18}
Na$_{2}$SO$_{4}$,\cite{wer:jctc10} and Li$_{2}$SO$_{4}$ \cite{plu:jpca13} at
concentrations below the solubility limit has been reported in simulation
studies.
Joung and Cheatham also provided some examples indicating that the anomalous
crystallization had lead in fact to erroneous interpretations of the results of
some previous computer simulations. It is thus important to ensure that the
force field has a solubility limit as close as possible to the experimental
value.

Unfortunately the calculation of the solubility by computer simulation is not a
trivial task. It started in 2002 with the pioneering contribution of Ferrario
et al.\cite{ferrario02} who calculated the solubility limit of KF in water. In
2007 Sanz and Vega,\cite{sanz:jcp07} evaluated the solubility of NaCl. Other
groups continued these calculations.\cite{paluch10,moucka11,ara:jcp12,mester15a,espinosa16}
After some initial discrepancies, a final agreement\cite{ben:jcp16,kol:jcp16,espinosa16}
was found for two common force fields for NaCl in water, namely the
Smith-Dang\cite{SmithDang} and the Joung-Cheatham models (both in SPC/E water).
The solubility was 0.7\,molal and 3.7\,molal, respectively, to be compared with
the experimental result of 6.1\,m. Since the nucleation of crystallites is
an activated process, the spontaneous precipitation is only observed at
concentrations 4-5 times larger than the solubility limit. This fact, which
could be viewed as an advantage, may also hide important inaccuracies of the
force field.

In 2010 Kim et al.\cite{kim:jpcb12} pointed out that none of the tested rigid,
non-polarizable models was able to reproduce the experimental trend for the
concentration dependence of the diffusion coefficient of water in electrolyte
solutions. Some of them even change the sign of the slope of the curves. They
concluded that the form of the interaction potentials had to be reexamined.
Similar conclusions were reached by Kann and Skinner.\cite{kan:jcp14}
Also recently it has been possible to determine activity coefficients for salts
in water.\cite{weerasinghe03} It has been shown that
the activity coefficient of salt increases too quickly with concentration. Due
to the Gibbs-Duhem relation, this also means that the activity of water
decreases too quickly ---as compared to experiment--- with the concentration of
the salt. Since the activity of water controls all the colligative properties,
one can not expect a good description of properties like cryoscopic descent,
osmotic pressure and vapor pressures of salts solutions.

After these facts the scenario looks rather depressive. Moreover, if the
situation of monovalent electrolytes is already bad, that of divalent ions as
Mg$^{2+}$, Ca$^{2+}$, SO$_{4}^{2-}$ is even worse. Everything seems to point
out that the inclusion of polarization is needed to describe ions in water.
Notice that developing a good polarizable force field is not an easy task.
If not parameterized with care sometimes a bad polarizable force field yields
worse results than a properly optimized non-polarizable one.
Since 2009, Leontyev and Stuchebrukhov\cite{leontyev09,leontyev10a,
leontyev10b,leontyev11,leontyev12,leontyev14} published a series of papers that
brought the last opportunity to non-polarizable force fields for ionic aqueous
solutions. They proposed that maybe the charge of the ions should not be an
integer number (in electron units), and that a scaled value should be used.
They argued that non-polarizable models do not fully account for the electronic
contribution to the dielectric constant and proposed that the screening
effect of the electronic continuum could be effectively included by a simple
scaling of the charges, namely, $q_{scaled}=q/\sqrt{\varepsilon_{el}}$ where
$\varepsilon_{el}$ is the high frequency dielectric constant of water. This
would led to a scaled charge for monovalent ions of about $0.75$. The use of
scaled charges may find its justification as coming from the charge transfer
and/or polarization of the water molecules around an ion.\cite{yao:jcp15}

The first paper of Leontyev and Stuchebrukhov in this line was not received with enthusiasm at the
beginning. After all, for a chemist is difficult to accept scaled charges since
the unscaled values have provided excellent results for ionic crystals.
 Jungwirth was probably the first to
advocate this idea.\cite{plu:jpca13,koh:jcpb14,koh:jpcb15,dub:jcpb17,mar:jcp18}
Kann and Skinner\cite{kan:jcp14} revisited the challenge of Kim et al.\cite{kim:jpcb12} and
concluded that the scaling of the charges improved the description of the
dynamical properties. Similarly, Yao et al.\cite{yao:jcp15} have shown that the
inclusion of dynamical charge transfer among water molecules accounts for the
distinct behavior of the water diffusivity in NaCl(aq) and KCl(aq). In the same
spirit, we have recently argued that the charges used in computer simulations
describe the potential energy surface rather than the dipole moment surface,
\cite{vegamp15} an idea that has been further expanded by Jorge and
coworkers.\cite{jor:jcp19,milne18} Even ourselves, after one year of efforts trying to
optimize a model for NaCl using unit charges for the ions, adopted the proposal
of Leontyev and Stuchebrukhov because of a simple argument: it seems to work.
We thus proposed a model for sodium chloride solutions, denoted as the Madrid
model,\cite{ben:jcp17} using $q_{scaled}=0.85$.  Recently, in a study of the
surface of a NaCl solution in water, \v{S}kv\'ara and Nezbeda have shown that
the results of the non-polarizable Madrid model and the polarizable AH/BK3
model\cite{kiss:jcp14} are found in most cases in mutual
agreement.\cite{skvara19}
Other work implementing the idea of charge scaling for electrolytes in
water is that of Fuentes-Azcatl and Barbosa\cite{fue:jpc16} who proposed a
force field for NaCl using $q_{scaled}=0.885$. Li and
Wang\cite{li:jcp15,li:jcpb17} have also applied the charge scaling concept
to monovalent ions using $q_{scaled}=0.804$. In summary, the number of groups
following the suggestion of Leontyev and Stuchebrukhov is growing in recent years
(see also Ref. \onlinecite{nico2018,cox_nacl})

Although it is clear that the description of ions in water may improve when
polarization is included, it seems of interest to probe non-polarizable force
fields using scaled charges. It is obvious that the scaling will have some
limitations but it may be convenient to explore its limits before going to the
polarizable models. The goal of this work is to develop a force field for ionic
solutions in water based on the use of scaled charges for the ions.
Since the performance of a force field for ions in water is related to the
performance of the model chosen to represent the water interactions, it is
essential to employ a satisfactory force field for water. In such a case good
predictions of the solution properties will also be obtained at the infinite
dilution limit. Among the rigid non-polarizable water models we have chosen
TIP4P/2005, which provides an excellent description of a number of properties
of liquid and solid water.\cite{vega09,vega11}
For monovalent ions we use a scaled value of the charge, in particular,
0.85 (in electron units). This was our optimized value of the ionic charges in
a recently developed model for NaCl in TIP4P/2005 water which is able to
describe the NaCl aqueous solution quite accurately.

Here we extend the idea to other ions. The species selected in this work are
essentially those typically found in seawater and biological
fluids. In particular we will consider the monovalent cations Na$^{+}$ and
K$^{+}$ (as well as Li$^{+}$) and the divalent cations Mg$^{2+}$ and Ca$^{2+}$.
For consistency, the scaled charges would be 1.7 in the latter case. With
regard to anions, the most interesting ones seem to be Cl${^-}$ and
SO$_{4}^{2-}$. The proposed force field is transferable in the sense that the
water-ion and ion-ion interactions would be the same regardless of the
composition of the system. In other words, the Cl-Cl or Cl-water interactions
would be the same in NaCl, KCl, MgCl$_{2}$ or CaCl$_{2}$ aqueous solutions so
that the force field could be used for any type of mixture.

The force field proposed in this work will be denoted as the Madrid-2019 force
field.  The target properties used to develop the force field were the
solution densities, radial distribution
functions, hydration numbers and densities related to those of the melt and of the solid (see discussion below). 
Solubility was not directly used as a target property. The main reason is that
it has become clear recently that the use of the scaling of the charges does
not lead to good estimates of the nucleation rate for salt
crystallization.\cite{pana2019} One should recognize from the very beginning
that the scaling prevents an accurate description of the solid phase and/or the
molten salts. The lack of transferability to other phases is the price to pay to improve the
description of the solutions. We have taken into account solubility in
an indirect way. We have checked that no spontaneous precipitation is observed
at the experimental solubility limit after long simulation runs in large systems.
We have also checked that the number of contact ion pairs (CIP) were always
relatively low as we have recently shown that these figures should not be too
high at the solubility limit.\cite{benavides17} Thus, even though we can not
guarantee that the proposed force field will reproduce the experimental values of
the solubility limits, we can at least guarantee that the system will not
precipitate at these concentrations, which otherwise would certainly
invalidate the outcome of the simulations. We shall present results not
only for pure salt solutions, but also for several ternary mixtures showing
that the predictions are also quite reasonable for these independent test systems. 


\section{ The Madrid 2019 force field }
%
%
We assume that the total energy of the system is given by the sum of the
potential energy between the molecules/ions of the system (pairwise
approximation). The interaction between any pair of atoms {\em i, j} of the
system is given by a coulombic term plus a Lennard-Jones (LJ) potential:
%
%
\begin{equation}
V(r_{ij})= \frac{1}{4\pi \varepsilon_{0}}\frac{q_{i}q_{j}}{r_{ij}} + 4\epsilon_{ij}\left[\left(\frac{\sigma_{ij}}{r_{ij}}\right)^{12}
- \left(\frac{\sigma_{ij}}{r_{ij}}\right)^{6} \right].
\end{equation}
Here, $\varepsilon_{0}$ is the vacuum permittivity, $q_{i}$, $q_{j}$ the
charges of atoms {\em i, j}, $\epsilon_{ij}$ the energy minimum of the LJ
potential and $\sigma_{ij}$ the LJ diameter. Water is described by the
TIP4P/2005 model.\cite{abascal05b}
This water model has a LJ interaction site at the oxygen and charges q$_{H}$,
and q$_{M}$ located respectively at the hydrogen positions and at a point M
placed near the oxygen along the H-O-H bisector (see Table \ref{tab_parameters}). 
For monovalent ions (Na$^+$, K$^+$, Li$^+$, Cl$^-$) the scaled value of the
charge is $|q_{scaled}|=0.85$ since that was the choice for the Madrid model
proposed previously.\cite{ben:jcp17} This choice was suggested
by Kann and Skinner\cite{kan:jcp14} and guarantees a good description of the
infinite dilution properties of monovalent ions in TIP4P/2005 water since it
compensates the low dielectric constant of the water model.  For consistency,
we assign a charge of 1.7 (in electron units) to monoatomic divalent cations
(Mg$^{2+}$, Ca$^{2+}$). Since we use scaled charges, the notation Na$^{+}$ 
(or Mg$^{2+}$, etc.) does not reflects the ionic charge. However, for 
simplicity, we respect throughout this paper the common ionic notation.

Modeling sulfate (a highly symmetric polyatomic anion) as a rigid body pose
some problems when one intends to carry out molecular dynamics using
constraints to preserve the molecular geometry. To avoid the inconveniences,
the sulfur atom can not have mass and should be treated as a dummy atom. Of
course, since the mass distribution does not affect the equilibrium
thermodynamic properties within classical statistical mechanics, the sulfur
mass can be distributed among the oxygens. In this way, the total
molecular mass is preserved, so most of the dynamical properties will not be
affected. It may slightly affect the rotation dynamics of the sulfate group.
In summary, we have modeled sulfate as a set of four
interacting sites at the positions of the oxygen atoms in a tetrahedral
arrangement around a charged massless sulfur. The experimental molecular weight
is then distributed among the four oxygens.
In accordance to the choice for other ions, the net charge of the sulfate group
is -1.7.  There are several ways of distributing the net charge between the
oxygen and sulfur atoms. We have found that a relatively wide range of values
for the sulfur charge (q$_S$) may account of the properties of sulfate
solutions although the parameters of the LJ interactions are slightly dependent
on the particular choice of q$_S$. We have finally assigned it a charge of 0.90
(in electron units).  The geometry of the sulfate molecules and the charges of
the sites used in this work are presented in Table \ref{tab_parameters}. 

\begin{table}[H]
\caption{Parameters of the Madrid-2019 force field. Charges of the particles used in this work and geometric
parameters of the water and sulfate molecules. Parameters for TIP4P/2005 water
were taken from Abascal and Vega. \cite{abascal05b} }
\label{tab_parameters}
  \begin{center}
    \begin{tabular}{ c c c c c c c c c c }
\hline
\hline
\multicolumn{10}{c}{Particle charges/$e$} \\
\hline
\multicolumn{10}{c}{$q_{_{Na}}=q_{_{K}}=q_{_{Li}}=0.85$, \,\,\,\,\, $q_{_{Mg}}=q_{_{Ca}}=1.70, \,\,\,\,\, $ $q_{_{Cl}}= -0.85$}\\
\multicolumn{10}{c}{$q_{_{S}}=0.90$, \,\,\,\,\, $q_{_{O_{s}}}= -0.65$, \,\,\,\,\,  $q_{_{H}} = -q_{_{M}}/2= 0.5564$}\\
\hline
\\
\multicolumn{5}{c}{H$_2$O geometry} & \multicolumn{5}{c}{SO$_4$ geometry} \\
\hline
\multicolumn{5}{c}{distance d$_{\text{OH}}=0.9572$~\AA} & \multicolumn{5}{c}{distance d$_{\text{OS}}=0.149$~\AA}\\
\multicolumn{5}{c}{distance d$_{\text{OM}}=0.1546$~\AA} & \multicolumn{5}{c}{distance d$_{\text{OO}}=0.243316$~\AA}\\
\multicolumn{5}{c}{Angle H-O-H=104.52$^\circ$}          & \multicolumn{5}{c}{Tetrahedral structure}\\
\hline
\hline
    \end{tabular}
  \end{center}
\end{table}

We proceed now to describe how the parameters of the LJ interactions have been
obtained.
The order in which the parameters of the force field were 
optimized was as follows:
\begin{itemize}
\item{Parameters of Na$^{+}$ and Cl$^{-}$.}
\item{Parameters of cations (using the parameters of Cl$^{-}$ obtained in the
previous step) from studies of the corresponding cation chlorides.}
\item{Parameters of SO$_{4}^{2-}$ from studies of Na$_{2}$SO$_{4}$.}
\item{Fine tuning of the K$^{+}$-SO$_{4}^{2-}$, Li$^{+}$-SO$_{4}^{2-}$ and Mg$^{2+}$-SO$_{4}^{2-}$ cross
interactions from studies of the corresponding sulfate solutions.}
\end{itemize}

The following set of target properties have been used to determine the optimal
LJ interactions parameters for each salt:
\begin{itemize}
\item{Densities of the aqueous solution at moderate (around 1\,molal) and high 
concentrations (close to the solubility limit).}
\item{Densities of the molten salt (at the experimental melting temperature)
and of the solid at ambient conditions  (only when the stable solid
had the rock salt structure).
For reasons that will be explained below, 
the target values for the densities of the molten salts and solids were not just the experimental ones.} 
\item{Position of the main peak of the ion-water radial distribution function and
hydration numbers.}
\item{ A moderate degree of clustering of ions near the experimental 
value of the solubility limit. In practice 
this is achieved by imposing a number of ionic pairs below 0.5  
at the experimental value of the solubility. 
This, indirectly guarantees that the solubility of the model is not too low when compared to 
experiments (see Ref.\onlinecite{benavides17} for a detailed discussion of this). Notice that experimental 
values of the number of CIP at the solubility limit are in general not available.}
\end{itemize}

In Table \ref{tab_melt_solub} the experimental melting temperatures\cite{haynes2014} and
solubility limits\cite{pen:cal03,boc:cjc61} for the salts considered in this work are presented. As it
can be seen, the solubility of all the salts (with the exception of CaSO$_{4}$)
are moderate or high. 
\begin{table}[H]
\caption{Experimental melting temperature for anhydrous salt\cite{haynes2014} and salt solubility in water at 25 $^{\circ}$C reported in molality units.\cite{pen:cal03,boc:cjc61}}
\label{tab_melt_solub}
  \begin{center}
    \begin{tabular}{ c c c}
\hline
\hline
Salt     & Melting     & Solubility\\
         & temperature & at 25 $^{\circ}$C \\
         & (K)         & (m) \\
\hline
LiCl             &  883.15 & 19.95 \\
NaCl             & 1073.85 & 6.15 \\
KCl              & 1044.15 & 4.81 \\
MgCl$_{2}$       &  987.15 & 5.81 \\
CaCl$_{2}$       & 1048.15 & 7.3 \\
Li$_{2}$SO$_{4}$ & 1132.15 & 3.12 \\
Na$_{2}$SO$_{4}$ & 1157.15 & 1.96 \\
K$_{2}$SO$_{4}$  & 1342.15 & 0.69 \\
MgSO$_{4}$       & 1397.15 & 3.07 \\
CaSO$_{4}$       & 1733.15 & 0.02 \\
\hline
\hline
    \end{tabular}
  \end{center}
\end{table}

Let us now expand a little bit our discussion on the target
properties.
The strongest forces in ionic systems come from electrostatic interactions, more
in particular from the potential between unlike-charged particles which are
somehow balanced by the repulsive forces at short distances. When
ionic crystals are dissolved, the ions fall quite apart and the ionic
interactions are much weaker and are replaced by the ion-water interactions.
In both systems the interactions between non-identical particles (cation-anion,
ion-water) are dominant. This is in contrast with typical choices of force
fields for non-ionic systems which usually define the parameters for the 
interactions between particles of the same type and evaluate the cross
interactions using some (arbitrary) rules. Apart of the use of scaled
charges, one of the main ideas behind this work is that the parameters
associated with the cation-anion and ion-water interactions should be
explicitly optimized. Since the charges of the particles are fixed, the
optimization process mostly affects the $\sigma_{ij}$ and $\epsilon_{ij}$ (with
$i\neq j$) parameters of the Lennard-Jones potential. In this way, the density
of the solution is paramount to determine the ion-water potential (notice that,
given the excellent performance of TIP4P/2005, the departures of the solution
densities from that of pure water provides a direct measure of the ion-water
interactions). We have thus optimized the LJ parameters of the ion-water
interactions in order to reproduce the density of moderate to high concentrated
solutions.

 On the other hand, in order to optimize the anion-cation parameters, we have
used information from the experimental densities of molten salts and solid phase.
When developing the Madrid model for aqueous NaCl, we
also considered the solid phase in the optimization process.  However, here we
have changed the approach in a qualitative way. Now we do not attempt to
predict the exact experimental values of the densities neither of the melt nor
of the solid since for most of the salts we failed to reproduce simultaneously the densities of
the melt and those of the solution at high concentrations. Thus we recognize that 
models with scaled charges should not reproduce
the experimental values of ionic systems without water (accurate predictions in this case 
could only be obtained when using
the full ionic charges). We observed that, typically, the density of the melt/solid 
 increases by about 20/8 per cent when the charge of the ions changes from 0.85 to 1 while
keeping the LJ parameters.  Thus, our target value for the density of the melt/solid
                 was around 20/8 per cent smaller than the experimental value. Besides 
this allows the possibility of developing in the future a 
polarizable version of the force field described here where the charge of the ions changes with the 
environment and return to the full charge values in the melt and/or solid phase.

As commented above cation-cation and/or anion-anion interactions have a minor
impact in the final force field and we have usually accepted the values of
previous works.
In some cases (mostly for the interactions between ions of different type but
carrying the same charge sign) we have used the Lorentz-Berthelot (LB)
combining rules
\begin{equation}
\sigma_{ij}=\frac{\sigma_{ii}+\sigma_{jj}}{2},  \,\,\, \epsilon_{ij}=\sqrt{\epsilon_{ii}\epsilon_{jj}}.
\label{eqLB}
\end{equation}
After these steeps, we had a preliminary set of parameters which allowed us to
compute the radial distribution functions of the aqueous solutions and calculate
the position of the ion-water first peak, the CIP at high concentrations and
the hydration numbers to check if the overall results were reasonable. These
calculations provided a feedback so the final refined parameters are the
outcome of some trial and error methodology.

Our starting point was the recently proposed Madrid model, a force field for
NaCl in water. However it was clear from the beginning that the densities
predicted for MgCl$_{2}$ and CaCl$_{2}$ were slightly lower than the
experimental ones. Since, in those attempts, we tried to optimize the
parameters for Mg$^{2+}$ and Ca$^{2+}$ while keeping the parameters of Cl$^{-}$ from the Madrid
model that was suggesting that the contact distance between Cl$^{-}$ and water was
too large. We thus reduced the parameter $\sigma_{Cl-Ow}$ from the
value in the Madrid model (0.426867) to 0.423867~nm. This is certainly a small
change but, to keep the excellent predictions of Madrid model, it was necessary
to slightly modify the rest of parameters. This new force field for aqueous
NaCl is denoted as Madrid-2019 model and can be regarded as a minor
modification of the original Madrid model (a comparison of the
parameters of both models is given in the Supplementary material). In the
results section we shall show that the performance of both models are
essentially the same.

The parameters of the Madrid-2019 force field are presented in two separated
Tables. The values of $\sigma$ for the LJ interactions are given in
Table~\ref{tab_sig} while the $\epsilon$ parameters are shown in Table
\ref{tab_eps} (some $\sigma_{ii}$ and $\epsilon_{ii}$ parameters
were taken from the literature: Li\cite{joung08}, Na\cite{ben:jcp17},
Mg\cite{elf:epjst16}, Ca\cite{mar:jcp18}, O$_w$\cite{abascal05b} and
S\cite{cannon94}).  The geometry and charges were presented previously in Table
\ref{tab_parameters}. 
When for a certain interaction one reads LB, it means that the corresponding
value was obtained from the application of the LB combining rule,
Eq.~(\ref{eqLB}). When a + symbol is written in addition to the word LB, then it
means that the LB rule has been validated in this work by the results of at
least one mixture.  
\begin{table}[H]
\caption{Parameters of the Madrid-2019 force field. Lennard-Jones $\sigma_{ij}$ parameters (in nm) for electrolyte
solutions in TIP4P/2005 water containing Li$^{+}$, Na$^{+}$, K$^{+}$,
Mg$^{2+}$, Ca$^{2+}$, Cl$^{-}$ and SO$_{4}^{2-}$. Some $\sigma_{ii}$ of cations
were taken from the literature (Li\cite{joung08}, Na\cite{ben:jcp17},
Mg\cite{elf:epjst16}, Ca\cite{mar:jcp18}, O$_w$\cite{abascal05b} and
S\cite{cannon94}). In cases where a numerical value is not given, we suggest to
follow Lorentz-Berthelot (LB) combination rules. LB(+) indicates that, in these
cases, we have checked LB combining rules in binary or ternary solutions with
satisfactory results. O$_{w}$ and O$_{s}$ denote the oxygen site in water and
sulfate, respectively.}
\label{tab_sig}
  \begin{center}
    \begin{tabular}{ c c c c c c c c c c }
\hline
\hline
       & Li     & Na     & K      & Mg     & Ca     & Cl     &O$_{w}$ & S      &O$_{s}$  \\
\hline
Li     &0.143970& LB(+)  & LB     & LB     & LB     &0.270000&0.212000& LB(+)  &0.284485 \\

Na     &        &0.221737& LB(+)  & LB(+)  & LB(+)  &0.300512&0.260838& LB(+)  & LB(+)   \\

K      &        &        &0.230140& LB     & LB     &0.339700&0.289040& LB(+)  &0.320000 \\

Mg     &        &        &        &0.116290& LB     &0.300000&0.181000& LB(+)  &0.240645 \\

Ca     &        &        &        &        &0.266560&0.315000&0.240000& LB     & LB      \\

Cl     &        &        &        &        &        &0.469906&0.423867& LB(+)  & LB(+)   \\

O$_{w}$&        &        &        &        &        &        &0.315890& LB(+)  &0.340445 \\

S      &        &        &        &        &        &        &        &0.355000& LB      \\

O$_{s}$&        &       &         &        &        &        &        &        &0.365000 \\
\hline
\hline
    \end{tabular}
  \end{center}
\end{table}

\begin{table}[H]
\caption{Parameters of the Madrid-2019 force field. As in Table \ref{tab_sig}, but for Lennard-Jones $\epsilon_{ij}$ parameters (in kJ/mol).}
\label{tab_eps}
  \begin{center}
    \begin{tabular}{ c c c c c c c c c c }
\hline
\hline
       & Li      & Na       & K        & Mg       & Ca       & Cl       & O$_{w}$  & S        & O$_{s}$ \\
\hline
Li    & 0.435090 & LB(+)    & LB       & LB       & LB       & 1.282944 & 0.700650 & LB(+)    & 0.803609 \\

Na    &          & 1.472356 & LB(+)    & LB(+)    & LB(+)    & 1.438894 & 0.793388 & LB(+)    & LB(+)    \\

K     &          &          & 1.985740 & LB       & LB       & 1.400000 & 1.400430 & LB(+)    & LB (+)   \\

Mg    &          &          &          & 3.651900 & LB       & 3.000000 & 12.00000 & LB(+)    & 2.748743 \\

Ca    &          &          &          &          & 0.507200 & 1.000000 & 7.250000 & LB       & LB       \\

Cl    &          &          &          &          &          & 0.076923 & 0.061983 & LB(+)    & LB(+)    \\

O$_{w}$&         &          &          &          &          &          & 0.774908 & LB(+)    & 0.629000 \\

S     &          &          &          &          &          &          &          & 1.046700 & LB       \\

O$_{s}$&         &          &          &          &          &          &          &          & 0.837400 \\
\hline
\hline
    \end{tabular}
  \end{center}
\end{table}

\section{Simulation details}
NVT and NPT molecular dynamics (MD) simulations have been performed using the
GROMACS package.\cite{hess08} In all runs the time step was 2~fs. The
cut-off radii were fixed at 1~nm for both electrostatics and Lennard-Jones
interactions. Long range corrections to the LJ part of the potential energy
and pressure were included. Coulombic interactions were evaluated with the
smooth PME method.\cite{essmann95}  We have used the Nos\'e-Hoover
thermostat\cite{nose84,hoover85} with a relaxation time of 2~ps to keep a
constant temperature, except in the case of solid state simulations where
v-rescale algorithm \cite{bus:jcp07} has been selected. NPT simulations were
performed at 0.1~MPa (and in some cases at 200 MPa) with Parrinello-Rahman
pressure coupling\cite{parrinello81} with a relaxation time for the barostat of
2~ps.  The LINCS algorithm\cite{hess97,hess08b} has been used to constrain
water geometry for most of the systems. In the case of salts containing the
sulfate anion we used SHAKE \cite{ryc:jcp77} for both water and sulfate because
this algorithm is more efficient for the sulfate group. 

To calculate equilibrium properties like density or radial distribution
function for liquid systems, we have performed simulations of 50 ns for a
system containing 555 water molecules.  This choice is useful because 10 ions
of any type correspond to a 1\,molal solution of that ion.
The number of contact ion pairs (CIP) can be computed from the cation-anion RDF as
\begin{equation}
    n^{CIP}= 4\pi \rho_{\pm}\int_{0}^{r_{min}} g_{+-}(r)\;r^{2}\;dr,
\label{eq_cip}
\end{equation}
where $g_{+-}$ is the cation-anion RDF and $\rho_{\pm}$ is the number density
of cation or anions. The integral upper limit $r_{min}$ is the
position of the first minimum in the RDF (if any) which must be located at a
similar distance to that of the cation-O$_{w}$ RDF. It is good idea to plot
simultaneously the cation-anion and cation-O$_{w}$ RDFs to determine if one is
really evaluating the CIP or a solvent separated ion pair. 
To check for the absence of precipitation we typically performed a long run (of
about 40~ns) for a system containing $555\times 8=4440$ molecules of water 
and the corresponding number of ions to give the
experimental value of the solubility limit and checked for aggregation.

The transport properties as viscosities and/or diffusion
coefficient were also determined in a system having 4440 water molecules.
The methodology used to compute the viscosity is similar to that described in previous 
works.\cite{gonzalez10}
A previous NPT simulation is required to
calculate accurately the volume of the system. After that, an 
NVT simulation was performed lasting between 50~ns and 200~ns.
Throughout the run, the pressure tensor $P_{\alpha\beta}$ 
was calculated and saved on disk every 2 fs. 
Finally, the Green-Kubo formula was used:
\begin{equation}
\label{shear_GK}
 \eta = \frac {V}{kT} \int_{0}^{\infty} 
 \langle P_{\alpha\beta}(t_{0})\; P_{\alpha\beta}(t_{0}+t) \rangle_{t_{0}}\; dt.
\end{equation}
The actual upper limit of the integral should be much higher than that of pure
water. In fact, for some solutions, we have used values of the order of several
hundreds of picoseconds (for pure water the upper limit is usually of the order
of 10~ps).

The self-diffusion coefficient has been evaluated using the Einstein relation:  
\begin{equation}
\label{eq_diffus}
  D = \lim_{t \to \infty} \frac{1}{6 t}  \Big \langle [\Vec{r}_{i}(t)-\Vec{r}_{i}(t_{0})]^{2} \Big \rangle ,
\end{equation}
where $\Vec{r}_{i}(t)$ and $\Vec{r}_{i}(t_{0})$ are the position of the i$-th$
particle at time t and a certain origin of time t$_{0}$ and the $\langle
[r_{i}(t)-r_{i}(t_{0})]^{2}\rangle$ term is the mean square displacement (MSD).

The densities of the melt and of the solid (with the rock salt structure) were
obtained for systems containing 1000 ions. Simulations typically lasted
10~ns. For the melts the simulations were performed at 0.1~MPa and at the
melting temperature of the anhydrous salt (see Table \ref{tab_melt_solub} ).
The simulations of the solid were performed at 298.15~K and 0.1~MPa for
the salts exhibiting a NaCl solid structure: LiCl, NaCl and KCl.
All these solids were found to be mechanically stable after a 10~ns run. 

\section{Results}

\subsection{1:1 electrolyte solutions}

\subsubsection{Sodium chloride}

\begin{center} \begin{figure}[H] \centering
\includegraphics*[clip,scale=0.4]{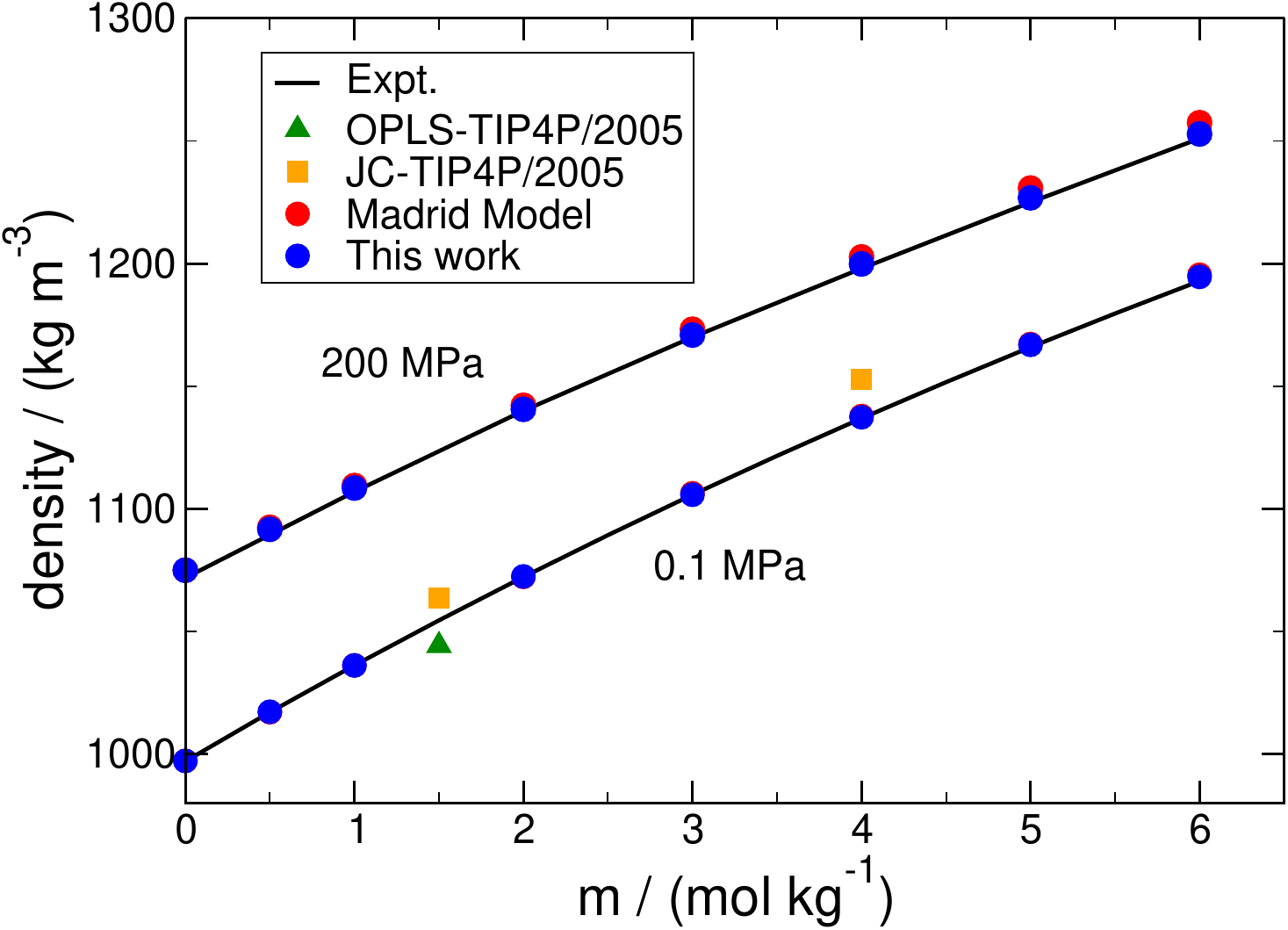}
\caption{
Density as a function of molality for aqueous NaCl solutions at $T=298.15$~K
for the 0.1 MPa and 200 MPa isobars. Values from this work are shown with blue
circles, the Madrid model predictions\cite{ben:jcp17} are the red circles, and
the results for the OPLS-TIP4P/2005\cite{opls} and
JC-TIP4P/2005\cite{ben:jcp17} force fields are represented by a green triangle
and orange square symbols, respectively. The experimental
data\cite{Pitzer_data,Pelper,2kbar} are shown as a continuous line.}
\label{fig_den_1_2000_bars}
\end{figure}
\end{center}

As mentioned above, we realized soon in this research that setting the
parameters of the chloride anion as in the original Madrid model did not allow
to reproduce with high accuracy the densities of MgCl$_{2}$ and CaCl$_{2}$
solutions. That was possible when we decreased by 0.003 nm the value of $\sigma_{Cl-O_{w}}$.
This slight change forced a fine tuning of the rest of parameters.
The new Madrid-2019 model can be regarded as a minor modification of the original one
(see the Supplementary Material which also includes a topol.top file of GROMACS with the 
parameters of the potential).
Fig.~\ref{fig_den_1_2000_bars} shows the density predictions using different
models for NaCl solutions. In the OPLS-TIP4P/2005,\cite{opls}
and JC-TIP4P/2005,\cite{ben:jcp17} models the ionic parameters are those of the
OPLS and Joung-Cheatham\cite{joung08}  force fields, respectively, and the water model is TIP4P/2005.
Since all these force fields use the same model for water, 
the differences in performance are due to differences the
ion-water and ion-ion interactions. As can be seen in the plot, the
calculations for the Madrid-2019 force field are in better agreement with the
experimental data\cite{Pitzer_data,Pelper,2kbar} than those of OPLS-TIP4P/2005 and JC-TIP4P/2005. The results
of this work are quite similar to those of the previous Madrid model although
we notice a slight improvement in the 200 MPa isobar at high concentrations.

In Table~\ref{tab_dens_melt_solid} densities for solid (at 0.1~MPa and 298.15~K)
and molten salts (at 0.1~MPa and the experimental melting temperature) are
presented. For each system two results are shown. The first one corresponds to
the density of the Madrid-2019 force field using the scaled charges
$q_{scaled}$. The second one shows the density obtained with the same set of
parameters as in the Madrid-2019 model but replacing the scaled charges by the
standard values $q$ (i.e 1 for monovalent and 2 for divalent ions). The
densities of the molten NaCl using the Madrid-2019 force field is a 15 percent
lower than in experiment. Concerning the density of the crystal the result of
our model is about 5 per cent below the experimental one.  These
departures greatly decrease when the scaled charges are replaced by the full
ones.  This fact seems to indicate that the relative size of the ions is well
captured by the parameters presented in Table \ref{tab_sig} and \ref{tab_eps}
for the ion-ion interactions.  In other words, the removal of water from the
solution translates into a change of the effective ionic charges from those
used in this work to the usual integer numbers in the crystal or the melt.  In
this way, the non-transferability of the force field between the solution and
the pure salt in condensed state is the price to pay for a better description
of the solution properties.  Notice finally that the results of
Table~\ref{tab_dens_melt_solid} indicate that the comments made here for NaCl
may be extended to other salts: the typical deviations between the scaled
charged model and the full charged ones are less than a 20 percent for the
melts and less than  8 percent for the crystal.

For the sulfate molecules we are not presenting results with the unscaled
charges as there are many ways of distributing the charge between the oxygen
and the sulfate to achieve a net charge of -2. 
\begin{table}[H]
\caption{Densities of the molten salts (at 0.1 MPa and the experimental melting
temperature\cite{haynes2014}) and of the anhydrous salt crystals (at 0.1 MPa
and 298.15 K). Values of this work are given under columns labelled q$_{sc}$
and q which are obtained using scaled charges and total charges for the
ions, respectively.}
\label{tab_dens_melt_solid}
  \begin{center}
    \begin{tabular}{c c c c c c c c}
\hline
\hline
Salt  & \multicolumn{3}{c}{Melt density} &\, & \multicolumn{3}{c}{Solid density} \\
\cline{2-4}
\cline{6-8}
      & q$_{sc}$ & q & Expt & \, & q$_{sc}$ & q & Expt \\
\hline

LiCl        & 1237 & 1443 & 1502 & & 1907 & 2075 & 2068 \\
NaCl        & 1331 & 1634 & 1556 & & 2050 & 2218 & 2165 \\
KCl         & 1236 & 1531 & 1527 & & 1834 & 1984 & 1984 \\
MgCl$_{2}$  & 1581 & 1779 & 1680   \\
CaCl$_{2}$  & 1754 & 2083 & 2085   \\
\hline
\hline
    \end{tabular}
  \end{center}
\end{table}

To gain further evidence of the quality of the Madrid-2019 force field we have analyzed some 
structural results. We have computed the ion-ion and ion-water radial distribution functions
at 6\,m, a concentration close to the experimental value of the solubility limit.
From the RDF it is possible to evaluate the number of contact ion pairs (see 
Eq.~(\ref{eq_cip})) and the hydration numbers (i.e the number of water molecules
around each ion). The latter are computed as in Eq.~\eqref{eq_cip} but
replacing $\rho_{\pm}$ by $\rho_{w}$ and $g_{+-}(r)$ by g$_{ion-O_{w}}(r)$
instead. 
The results are summarized in Table \ref{tab_res_1_1_elect} and compared with
experimental X-ray and neutron diffraction data collected in the work of
Marcus.\cite{mar:cr88} The number of CIP is an important property when studying
electrolytes. A high value provides an indirect indication of cluster formation
and/or precipitation of the salt.  For 1:1 electrolytes with solubility lower
than 10\,m, Benavides et al.\cite{benavides17} suggested that the number of CIP
must be below 0.5 to be sure that precipitation and/or aggregation of ions has
not occurred. For NaCl we have found a value of 0.17 at 6\,m. In fact
aggregation was not found at this concentration in long runs of large systems.
The hydration numbers for Na$^{+}$ and Cl$^{-}$ are 5.4 and 5.9 waters,
respectively.  These values are within the range reported in
experiments.\cite{mar:cr88} The d$_{Na-O_{w}}$ and d$_{Cl-O_{w}}$ distances
are, 2.25~\AA\ and 3.10~\AA, respectively, to be compared to the experimental
estimations 2.33~\AA, and 3.05~\AA. 
\begin{table}[H]
\caption{Structural properties for 1:1 electrolyte solutions at 298.15 K and
0.1~MPa.  Number of contact ions pairs (CIP), hydration number of cations
(HN$_{c}$) and anions (HN$_{a}$), and position of the first maximum of the
cation-water (d$_{c-O_{w}}$), and anion-water (d$_{a-O_{w}}$) RDFs. In
parentheses, experimental data taken from the work of Marcus\cite{mar:cr88} and
references therein. Properties were calculated for LiCl, NaCl, and KCl
solutions at 12\,m, 6\,m, and 4.5\,m concentration, respectively.}
\label{tab_res_1_1_elect}
  \begin{center}
    \begin{tabular}{ c c c c c c c c}
\hline
\hline
Salt    & CIP   & HN$_{c}$    & HN$_{a}$   & d$_{c-O_{w}}$ (\AA)& d$_{a-O_{w}}$ (\AA)\\
\hline
LiCl    & 0.01  & 4(3.3-5.3)   & 5.9(4-7.3)   & 1.84(1.90-2.25) & 3.03(3.08-3.34) \\
NaCl    & 0.17  & 5.4(4-8)     & 5.9(5.5-6)   & 2.33(2.41-2.50) & 3.05(3.08-3.20) \\
KCl     & 0.38  & 6.5(6-8)     & 5.8(6-8)     & 2.73(2.60-2.80) & 3.03(3.08-3.16) \\
\hline
\hline
    \end{tabular}
  \end{center}
\end{table}

\begin{center}
\begin{figure}[H]
\centering
\includegraphics*[clip,scale=0.4]{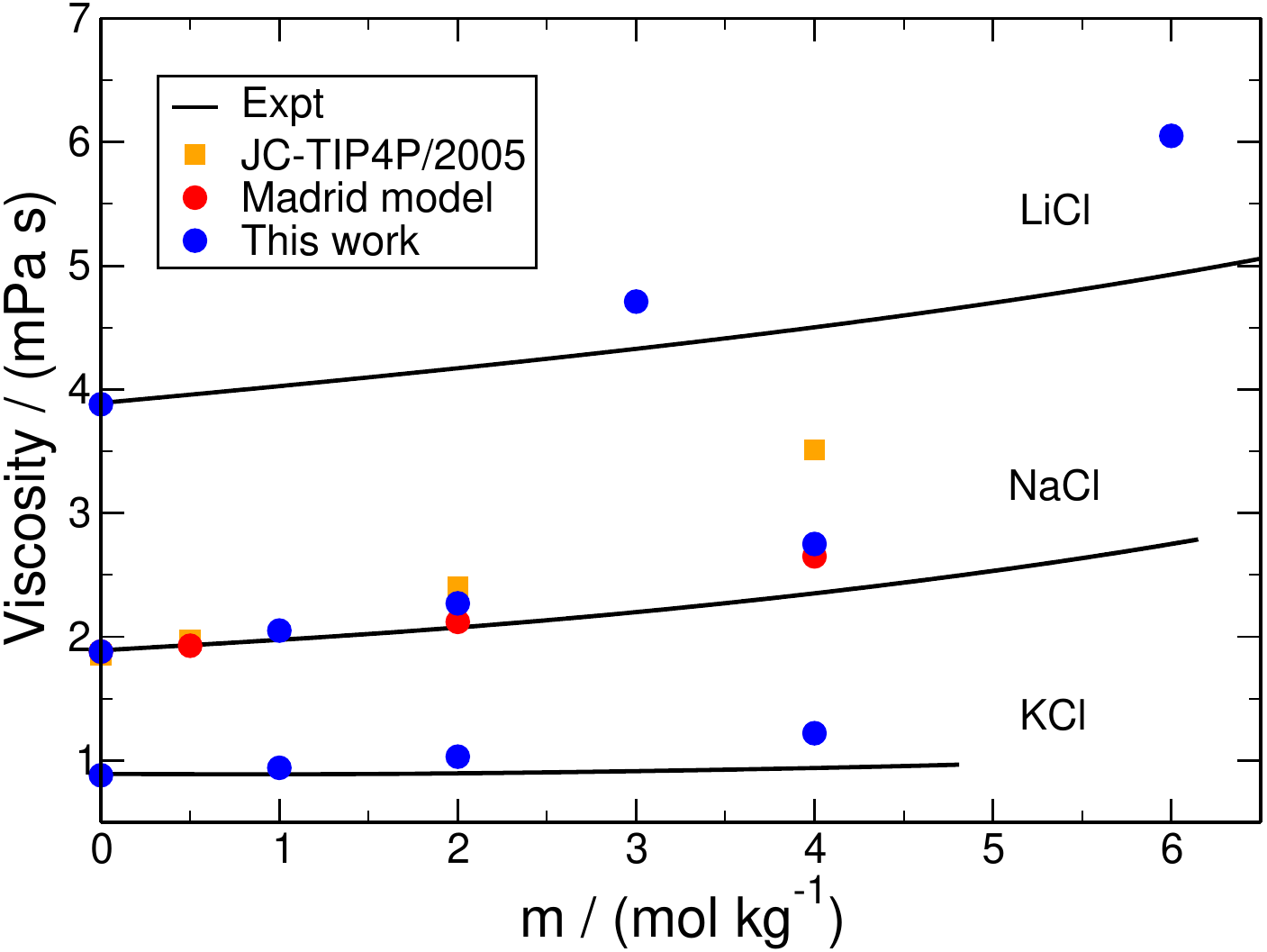}
\caption{Shear viscosity  curves as a function of concentration for aqueous
LiCl, NaCl, and KCl systems at 298.15 K and 0.1 MPa.
Result from this work are shown with blue circles, those from Madrid
model\cite{ben:jcp17} with red circles, JC-TIP4P/2005\cite{ben:jcp16} in orange
squares.
The continuous line is our fit of experimental data taken from
Refs.~\onlinecite{lal:jced07,lal:jced09} and references therein.
NaCl and LiCl values were shifted up one and three
units, respectively, for better legibility.}
\label{fig_visco_licl_nacl_kcl}
\end{figure}
\end{center}

Two transport properties have been evaluated in this work: the shear viscosity
and the self-diffusion coefficients. Fig.~\ref{fig_visco_licl_nacl_kcl} shows
the shear viscosity for NaCl aqueous systems as a function of the concentration
at 298.15 K and 0.1 MPa. The results of the Madrid-2019 force field are again
quite similar to those of the Madrid force field. In general, the agreement with
experimental data\cite{lal:jced07,lal:jced09} is excellent at low concentrations. The dependence with 
temperature follows the same trend as the experimental one but the slope is somewhat
overestimated, especially at very high concentrations. Anyway, our model improves
considerably the results for JC-TIP4P/2005. This confirms that the scaling of
the charges leads to a better description of the viscosity of NaCl(aq).

Numerical results for the self-diffusion coefficients of ions and water are
presented in Table~\ref{tab_diffusion}. The cation and anion self-diffusion
coefficients were evaluated at two different concentrations (0.5\,m
and 1\,m) to provide an idea of the value at low concentrations so that they can
be compared to the experimental ones at infinite dilution.\cite{li:gca74}
The diffusivity of Na$^{+}$ seems to be in accordance with the experimental measurements but
that of Cl$^{-}$ differs to a certain extent from experiment.
Finally, the performance of the force field for the ratio $D^{salt}_{w}/D^{pure}_{w}$ is
quite good in comparison with experimental data from M\"uller and Hertz.\cite{mul:jpc96}
For this ratio, we have not applying the finite-size correction to the water diffusion proposed by Yeh and Hummer\cite{yeh04}.
\begin{table}[H]
\caption{Self-diffusion coefficients of cations and anions ($D_c$, $D_a$) as a
function of concentration (in 10$^{-5}$ cm$^{2}$/s units) for electrolyte water
solutions at 298.15 K and 0.1 MPa. Comparison is done with experimental results
at infinite dilution.\cite{li:gca74}
Water diffusion in electrolyte solutions relative to water diffusion in pure
water, $D^{salt}_{w}/D^{pure}_{w}$, is also shown. Experimental data at 1\,m
were taken from M\"uller and Hertz,\cite{mul:jpc96} where $D^{pure}_{w}=2.3$
and in our simulation $D^{pure}_{_{TIP4P/2005}}=2.14$ without applying the 
finite-size correction of Yeh and Hummer.\cite{yeh04}
}
\label{tab_diffusion}
  \begin{center}
    \begin{tabular}{ c c c c c c c c c c c}
\hline
\hline
Salt & \multicolumn{3}{c}{$D_{c}$} & \, &\multicolumn{3}{c}{$D_{a}$} &\, & \multicolumn{2}{c}{$D^{salt}_{w}/D^{pure}_{w}$} \\
\cline{2-4}\cline{6-8}\cline{10-11}
    & \multicolumn{2}{c}{Sim}     & Expt & \, & \multicolumn{2}{c}{Sim}  & Expt & \, &Sim   & Expt \\
\cline{2-3} \cline{6-7} \cline{10-10}
    & 1\,m   & 0.5\,m     & 0 m  & \, & 1\,m   & 0.5\,m     & 0 m & \, & 1 m   & 1 m \\
\hline
LiCl  & 0.81 & 0.92 & 1.03 & \,& 1.31 & 1.38 & 2.03 &\,& 0.84 &  0.91\\
NaCl  & 0.99 & 1.22 & 1.33 & \,& 1.33 & 1.39 & 2.03     &\,& 0.89 &  0.94\\
KCl   & 1.63 & 1.70 & 1.96 & \,& 1.52 & 1.45 & 2.03     &\,& 0.97 &  1.03\\
MgCl$_{2}$ & 0.41 & 0.64 & 0.705 & \,& 0.89 & 1.30 &2.03 &\,& 0.67 & 0.73  \\
CaCl$_{2}$ & 0.55 & 0.63 & 0.791 & \,& 1.10 & 1.23 &2.03 &\,& 0.74 & 0.82 \\
Li$_{2}$SO$_{4}$& 0.63 & 0.82 & 1.03 & \, & 0.68 & 0.72 & 1.07 & \, & 0.74  &\\
Na$_{2}$SO$_{4}$ & 0.82 & 0.85 & 1.33 & \, & 0.76 & 0.91 &1.07 & \, & 0.79  & \\
K$_{2}$SO$_{4}$& -- & 1.60 & 1.96 & \, & -- & 0.86 &1.07 & \, & --   &\\
MgSO$_{4}$ & 0.40 & 0.59 & 0.705 & \, & 0.58 & 0.78 &1.07 & \, & 0.72 &\\
\hline
\hline
    \end{tabular}
  \end{center}
\end{table}

In summary, the main conclusion of the calculations for NaCl solutions is that
the Madrid-2019 model yields quite similar results to those obtained with the
original Madrid model for equilibrium, structural and transport properties. We
thus believe that other properties non evaluated in this work will follow a
similar pattern (for a more comprehensive discussion of the performance of the
Madrid model we refer the reader to our previous work\cite{ben:jcp17}).

\begin{center}
\begin{figure}[H]
\centering
\includegraphics*[clip,scale=0.4]{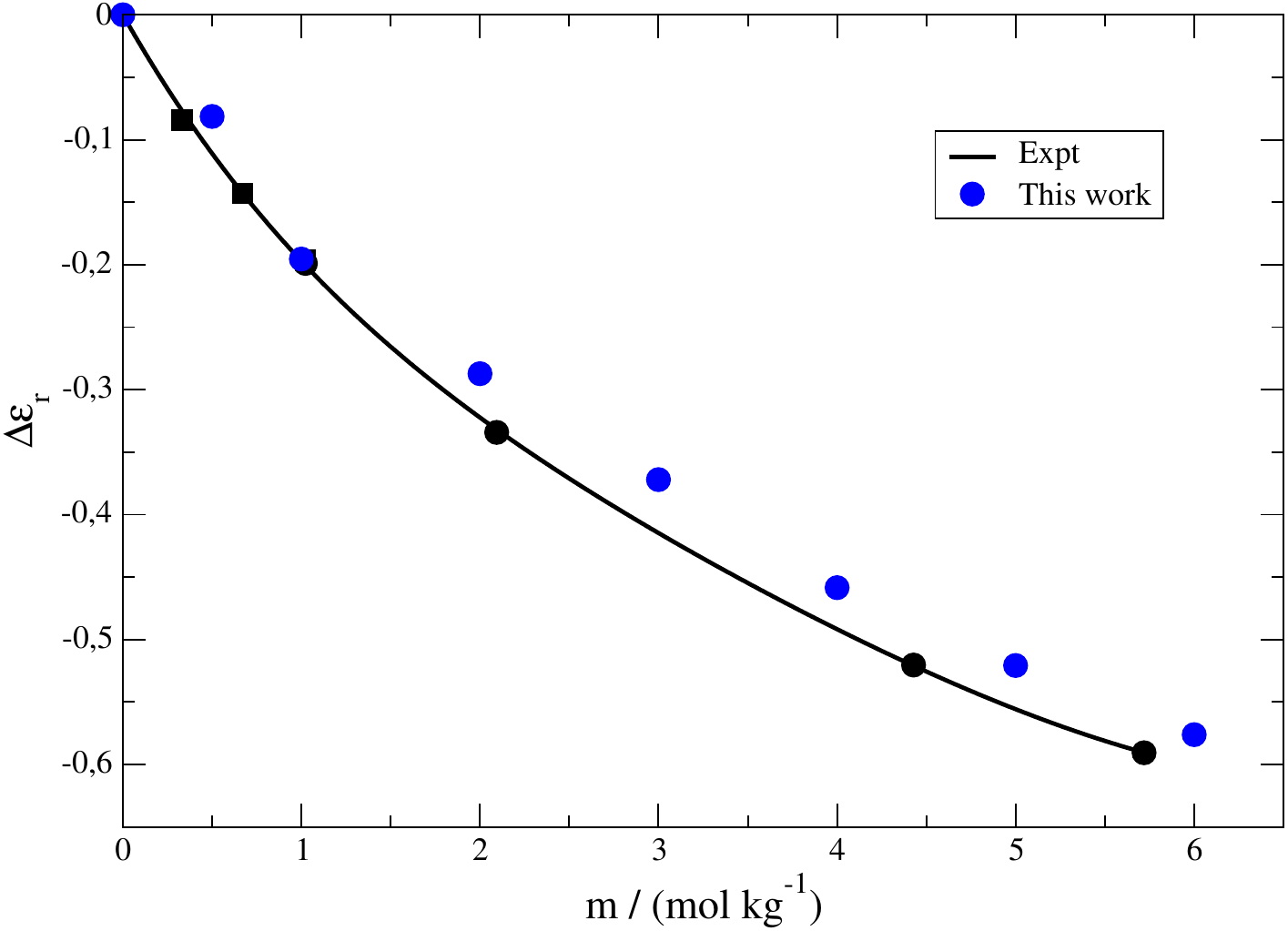}
\caption{Relative change of the dielectric constant as a function of the
salt molality for NaCl aqueous solutions at 1 bar and 298.15 K. Results using the Madrid-2019 force field are shown in blue circles, experimental values are also shown with a fitted curve of the data (black squares\cite{haggis_dielectric} and
black circles\cite{chris_dielectric}).}
\label{fig_dielectric_nacl}
\end{figure}
\end{center}

Fig. \ref{fig_dielectric_nacl} shows the behavior of the relative change of the dielectric
constant, $\Delta \varepsilon_{r} = (\varepsilon_{solution} -\varepsilon_{H_2O}) / \varepsilon_{H_2O}$,
of the NaCl solutions at room temperature. Experimental results were taken from
Refs.\onlinecite{haggis_dielectric,chris_dielectric}. As it can be seen,
the relative decrease of the dielectric constant with the salt concentration
is captured quite well by the Madrid-2019 force field  even though
the absolute values of the dielectric constant are not well described by the force field.
The reason for that is that the TIP4P/2005  model of water does not reproduce well the experimental
value of the dielectric constant of water (although the reason for that and possible ways to fix it has been discussed
in detail in previous work\cite{vegamp15,jor:jcp19} ).

\subsubsection{Potassium chloride}
\begin{center}
\begin{figure}[H]
\centering
\includegraphics*[clip,scale=0.4]{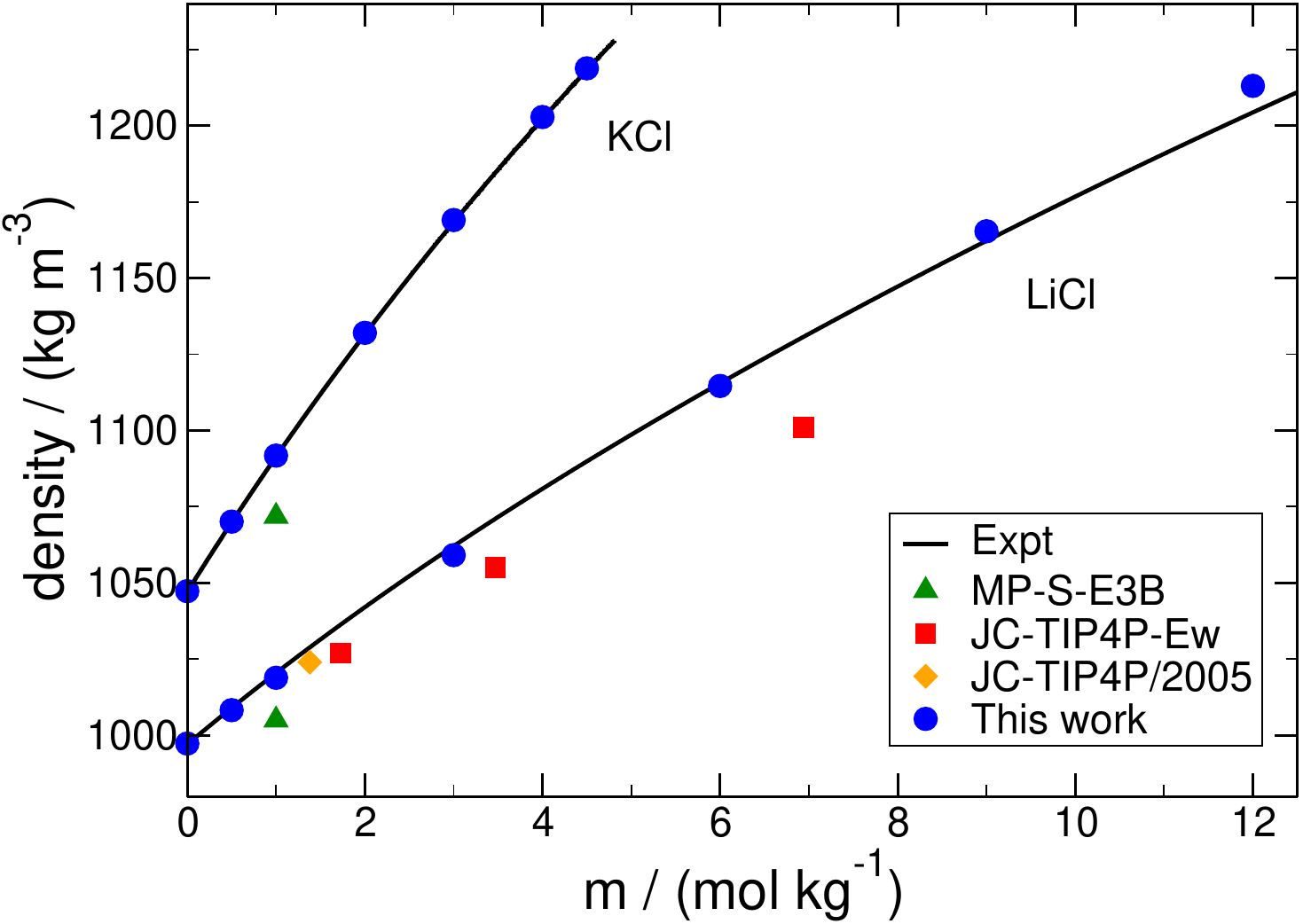}
\caption{Density as a function of molality at T= 298.15 K and 0.1 MPa  for two
1:1 electrolyte aqueous solutions, LiCl and KCl. The symbols are as follows.
Blue circles: this work, red squares: JC-TIP4P-Ew,\cite{joung08} orange
diamonds: JC-TIP4P/2005,\cite{ara:jpcb14} green triangles: MP-S/E3B.\cite{kan:jcp14}
The continuous line is our fit of experimental data taken from
Refs.~\onlinecite{lal:jced04, lal:jced07} and references therein.
KCl values were shifted up 50 units for better legibility.}
\label{fig_dens_licl_kcl}
\end{figure}
\end{center}

The densities of potassium chloride solutions are shown in Fig.
\ref{fig_dens_licl_kcl}. The agreement with experimental data is excellent
along the whole concentration range. The relative failure of MP-S/E3B model
which also uses scaled charges\cite{kan:jcp14} is in probably due to the fact
that the model underestimates the water density. For a concentration of 4.5\,m,
the K-O$_{w}$ RDF shows a first peak at 2.73~\AA\ and the hydration number of K
is 6.5 water molecules. At this concentration the maximum in the Cl-O$_{w}$ RDF
is located at 3.03~\AA\ and the corresponding coordination number is 5.8 water
molecules, very similar to that in LiCl and NaCl solutions. As for NaCl, the
structural results are in accordance with the experimental estimations. At the
highest simulated concentration the CIP takes an acceptable value of 0.38 without
cluster formations during the whole simulation. In contrast, precipitation of
KCl has been reported in previous work when using models with full ionic
charges.\cite{auf:jctc07}

The viscosities of KCl solutions were already shown in Fig. \ref{fig_visco_licl_nacl_kcl}
together with the results for other 1:1 electrolytes.  The comparison with
experiment is quite remarkable, especially because the model correctly predicts
the small dependence of $\eta$ on concentration.
The diffusion coefficients of the cation at 1\,m and 0.5\,m (see Table~\ref{tab_diffusion})
seem to point toward the experimental value at infinite dilution. As to the
Cl$^{-}$ anion, despite that there is some improvement in the self-diffusion
respect to LiCl and NaCl cases, the results of the model are clearly lower than
those from experiment. 
On the other hand, the presence of KCl does not affect much the diffusion
coefficient of water (the relative water diffusion at 1\,m is 0.97). In
experiments the same is true, although the salt changes the diffusion
coefficient in the opposite direction as it increases to 1.03
(this tendency is correctly described by the MP-S/E3B model\cite{kan:jcp14}).

\subsubsection{Lithium chloride}
Fig.~\ref{fig_dens_licl_kcl} shows the equation of state for LiCl in water
298.15 K and 0.1~MPa compared to experiment and to the results of other force fields. 
The performance of the Madrid-2019 is better than that of the MP-S/E3B model which also use
scaled charges. 
Results for force fields that do not use the concept of charge scaling are also
presented, in particular, for JC/TIP4P-Ew,\cite{joung08} and its modified
version\cite{ara:jpcb14} which replaces the TIP4P-Ew water model by TIP4P/2005.
The predictions of the Madrid-2019 force field are in excellent agreement with
experimental data. Although slight deviations may be noticed at very high
concentrations, the departure from experiment at 12\,m is less than 1\%. In
contrast, the results of the rest of force fields are not accurate enough at low
concentrations and, in the case of JC/TIP4P-Ew, it predicts a much less steeper
slope.

Some structural results for a 12\,m concentration were presented in Table
\ref{tab_res_1_1_elect} .
The CIP are very low at this concentration. This is probably due
to a strong hydration layer which prevents close cation-anion contacts. 
The Li$^{+}$ ions are hydrated with four water molecules which is consistent 
with the experimental values.\cite{mar:cr88}
In line with other monovalent halides, the hydration number of Cl$^{-}$
is 5.9. The ion-water distances at the first maximum of the RDF seem a bit
underestimated, about 3\% below the experimental estimations.  More consistent
results were found for other models, see Table VIII in the work of Kann and
Skinner.\cite{kan:jcp14}

The Madrid-2019 model exaggerates the dependence of the viscosity with concentration
(see Fig. \ref{fig_visco_licl_nacl_kcl}).
This is a common trend for all 1:1 electrolytes although it seems that the
deviation between simulation and experiment becomes larger as the size of the
cation becomes smaller. 
It is to be noticed that the differences between simulation and experiment are 
much larger for models with full ionic charges.
Thus, the charge scaling alleviates but does not correct completely the dependence
of the viscosity on salt concentration.
Concerning the values of the diffusion coefficient of Li at infinite dilution,
the results obtained in this work seem reasonable as compared to the experimental
value at infinite dilution. 
Notice that the differences between the diffusion of the chloride anion in
NaCl, KCl and LiCl solutions are quite modest suggesting that the effect of the
counter ion is small at low concentrations.  
The predictions for the water diffusion agree reasonably with the experimental data, 
the discrepancy is about a 8\% (see Table~\ref{tab_diffusion}).

\begin{center}
\begin{figure}[H]
\centering
\includegraphics*[clip,scale=0.4]{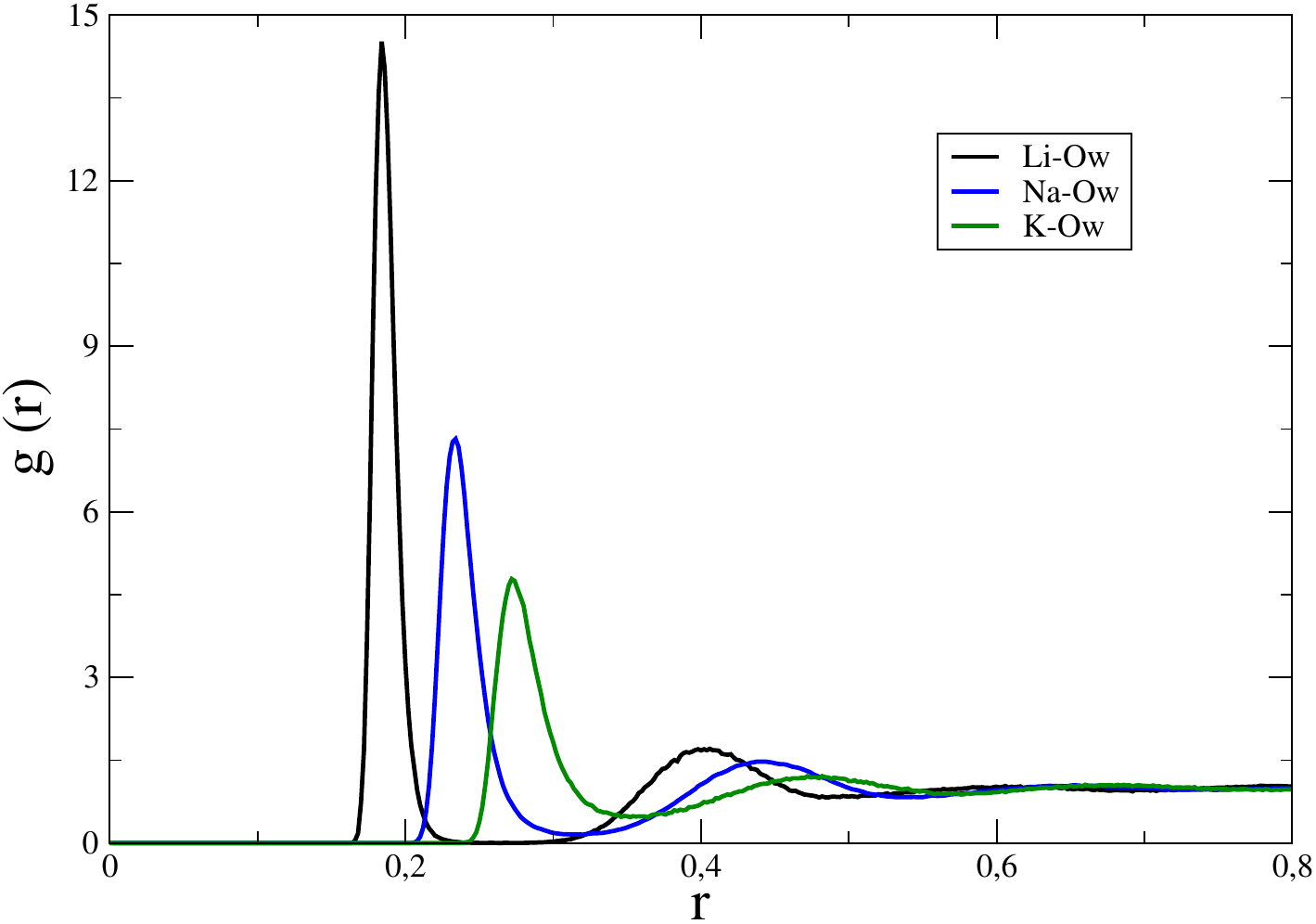}
\caption{Cation-water oxygen radial 
distribution function for 1:1 electrolyte solutions at 298.15K, 1 bar, and 1 molal as were obtained 
with the Madrid-2019 model in solutions: LiCl (black line), NaCl (blue line), and KCl (green line). }
\label{fig_rdf_li_na_k-ow}
\end{figure}
\end{center}

Finally we shall present the cation-oxygen radial distribution functions for LiCl, NaCl and KCl.
They are presented in Fig.\ref{fig_rdf_li_na_k-ow}. As it can be seen the obtained values reflects the size of the cations and the stronger hydration of the small cations.

\subsection{2:1 electrolyte solutions}

\subsubsection{Magnesium chloride}

\begin{center}
\begin{figure}[H]
\centering
\includegraphics*[clip,scale=0.4]{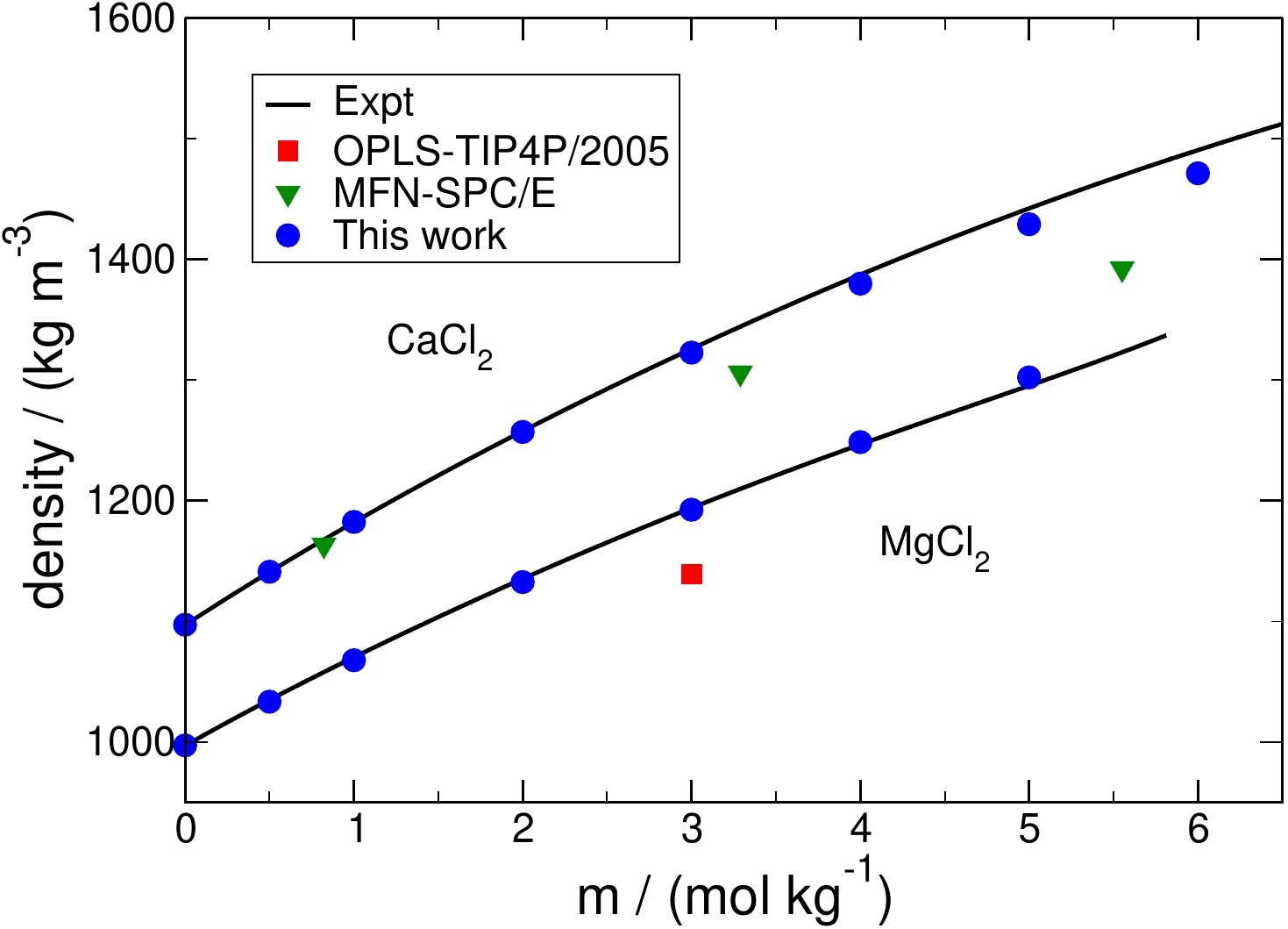}
\caption{Density curves for MgCl$_{2}$ and CaCl$_{2}$ in water solutions at 
298.15~K and 0.1~MPa. Blue circles are the data from this work, green triangles
are the  MFN-SPC/E results,\cite{mam:jcp13} and red squares represent the
calculations for the OPLS-TIP4P/2005 model.
The continuous line is our fit of experimental data taken from
Refs.~\onlinecite{lal:jced04, lal:jced09} and references therein. For a better appraisal, the
CaCl$_{2}$ densities have been shifted up 100~units.}
\label{fig_dens_mgcl2_cacl2}
\end{figure}
\end{center}

\begin{center}
\begin{figure}[H]
\centering
\includegraphics*[clip,scale=0.4]{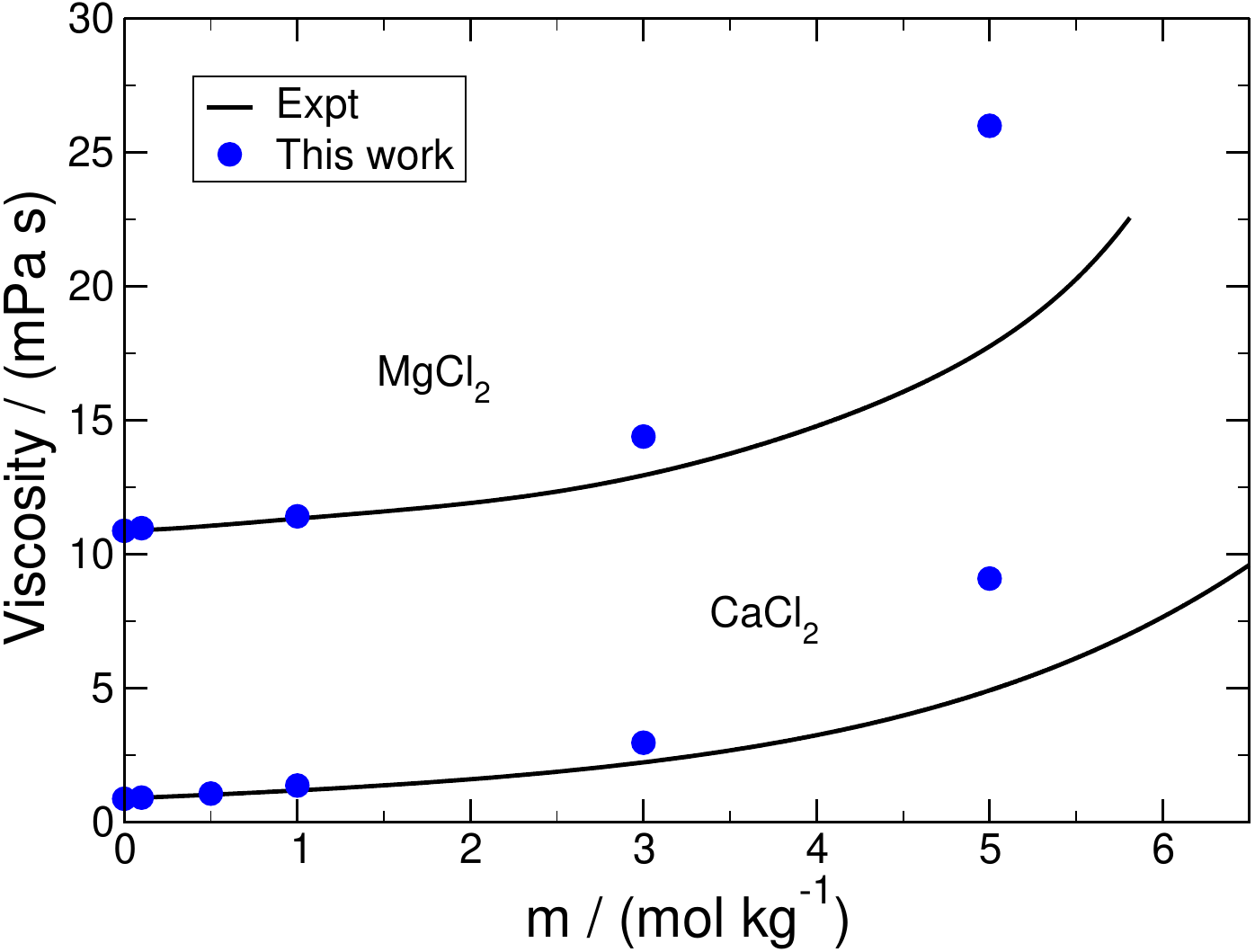}
\caption{Shear viscosity as a function of molality for aqueous MgCl$_{2}$ and
CaCl$_{2}$ solution at 298.15 K and 0.1 MPa. Result from this work are shown
with blue circles and the continuous line is our fit of experimental data taken
from Refs.~\onlinecite{lal:jced07, lal:jced09} and references therein.
MgCl$_{2}$ values were shifted up ten units.}
\label{fig_visco_mgcl2_cacl2}
\end{figure}
\end{center}

Fig.~\ref{fig_dens_mgcl2_cacl2} presents the densities of MgCl$_{2}$ solutions
at 298.15~K and 0.1~MPa. The agreement with the experimental data\cite{lal:jced04, lal:jced09} is quite good.
At high concentrations the deviations are of about 1 \%. The results obtained
with the OPLS force field in combination with TIP4P/2005 are also shown for a
concentration of 3\,m.  The deviation from experiment is quite noticeable for
this model that uses full ionic charges.
Structural properties at 5\,m and are summarized in Table
\ref{tab_res_2_1_elect}.  The hydration sphere around the  Mg$^{2+}$ contains
exactly six water molecules. No anions are able to enter within the hydration
shell of the cations (hence the CIP is 0) meaning that the interaction
between Mg$^{2+}$ and water is very strong. This is a well known result. Not
surprisingly MgCl$_2$ often precipitates as an hexahydrate.
The second hydration shell is found at 4.18~\AA\ with a coordination number of
13 waters in agreement with values reported in literature (see Table 3 in the
work of Zhang et al.\cite{zha:ejm19}).
The hydration number for Cl$^{-}$ is 5.9 in agreement with experiment. For both
cation and anion, the position of the first peak of the ion-O$_w$ RDF lie at a
shortest distance than the experimental estimations. This deserves a comment.
It is somewhat surprising that, even though the model is able to reproduce
extraordinary well the density of both pure water and the solution, the
distance of the ions to the oxygens in the first hydration shell is smaller
than that reported from scattering experiments.  It should be pointed out that
extracting individual RDF from scattering data is not a trivial task
(especially in electrolyte solutions where one has contributions from many
species). Even for pure water, experimental results for the RDF are improved
year after year.  For this reason at this point we believe that these
experimental data should be regarded with care. This problem might be revisited
in the future. In fact it would be of interest to compute the structure factor
from the simulations of this work and to see if they are able to describe the
experimental data.
\begin{table}[H]
\caption{As in Table \ref{tab_res_1_1_elect} but for 2:1 electrolyte
solutions. Properties in solution were calculated at 5\,m and 6\,m
for MgCl$_{2}$ and CaCl$_{2}$, respectively. CIP is just the number of anions in contact with the cation.}
\label{tab_res_2_1_elect}
  \begin{center}
    \begin{tabular}{ c c c c c c}
\hline
\hline
Salt        & CIP   & HN$_{c}$     &  HN$_{a}$   &d$_{c-O_{w}}$(\AA)& d$_{a-O_{w}}$(\AA)\\
\hline
MgCl$_{2}$  & 0     & 6(6-8.1)     & 5.9(6)       & 1.92(2.00-2.11) & 3.03(3.13-3.16)    \\
CaCl$_{2}$  & 0.02  & 7.1(5.5-8.2) & 6.2(5.8-8.2) & 2.38(2.39-2.46) & 3.03(3.12-3.25)    \\
\hline
\hline
    \end{tabular}
  \end{center}
\end{table}
Fig. \ref{fig_visco_mgcl2_cacl2} displays the concentration dependence of the
viscosity of MgCl$_{2}$ solutions. 
The Madrid-2019 model is able to describe the experimental behavior up to 3\,m
concentration but beyond this concentration the deviation from experiment is
significant. Since viscosities can be measured with great accuracy,
all the evidence presented so far points out that although the scaling of the
charge provides a reasonable description of transport properties up to
moderately high concentrations, the model fails at very high salt
(though much less than models that use full ionic charges). Something is going
on at high concentrations that is not captured by the force field of this work
(and probably for the rest of the force fields proposed so far). This is a
point that require further work. 
As to the diffusion of Mg$^{2+}$ (see Table~\ref{tab_diffusion}), our results
at 1\,m and 0.5\,m concentrations extrapolate nicely to the experimental data
at infinite dilution. The model also captures quite well the sizable decrease
in the diffusion coefficient of water when adding  MgCl$_{2}$. Experimentally
the diffusion coefficient of water in a 1\,m solution decreases to a 73\%,
to be compared to the prediction of the Madrid-2019 force field which yields
67\%.

\subsubsection{Calcium chloride}
The densities of calcium chloride solutions are displayed in
Fig.~\ref{fig_dens_mgcl2_cacl2}.
The performance of the Madrid-2019 model is excellent up to 4\,m. The
deviations from experiment increase slightly with the salt concentration but
they never exceed a 1\%.
Also included in the plot are the data reported by Mamatkulov et
al.\cite{mam:jcp13} here denoted as MFN-SPC/E. The discrepancies from
experiment are now important and exceed a 5 percent at high salt. The
hydration number of Ca$^{2+}$ is 7.1 water molecules (see Table
\ref{tab_res_2_1_elect}). The increase respect to that of Mg$^{2+}$ is brought
about by the increase in the cation size.  In fact, the maximum of the Ca-O$_w$
RDF appears at a higher distance, 2.38~\AA, than in the lighter divalent
cation. This distance is just at the lower limit of the range reported for
experiments.
The simulations of calcium chloride solution confirms that the hydration of the
chloride anion is essentially insensitive to the counterion. Only the
coordination number, 6.2, is marginally different.

The predictions for the viscosity of aqueous CaCl$_{2}$ follow a similar trend
to those for MgCl$_{2}$. The model does a good job up to 3\,m and fails
beyond this salt concentration (although the deviations from experiment are now
somewhat smaller). The diffusion coefficients of Ca$^{2+}$ in water seem to approach
the experimental value at infinite dilution (see Table \ref{tab_diffusion}).
The model also accounts quite acceptably for the experimental drop of the
diffusivity of the water molecules in a 1\,m solution (18 {\em vs} 26 per
cent).  It is worth noting that the force field proposed in this work captures
correctly the fact that, at 1\,m concentration, the diffusion of
water in 1:1 electrolytes decreases by about a 5 per cent whereas the change is
about a 20 per cent for 2:1 electrolyte solutions.

\begin{center}
\begin{figure}[H]
\centering
\includegraphics*[clip,scale=0.4]{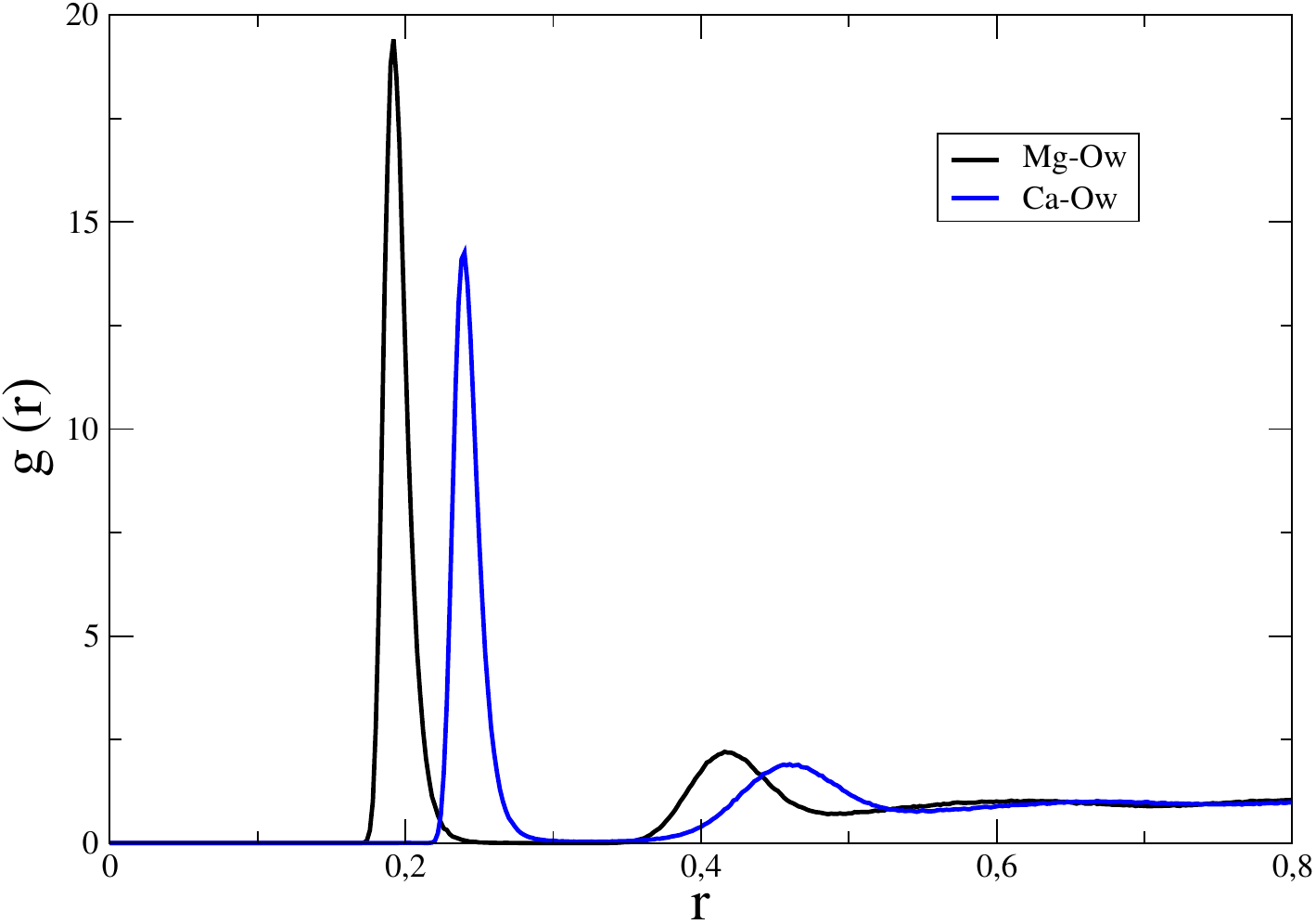}
\caption{Cation-water oxygen radial distribution function for 2:1 electrolyte solutions at 298.15 K, 1 bar, and molality 1 as were obtained
with the Madrid-2019 model in MgCl$_{2}$ solutions  (black line) and CaCl$_{2}$ solutions (blue line).}
\label{fig_rdf_mg_ca-ow}
\end{figure}
\end{center}

Finally we shall present the cation-oxygen radial distribution functions for MgCl$_{2}$ and CaCl$_{2}$ solutions.
Results are presented in Fig.\ref{fig_rdf_mg_ca-ow}. As it can be seen the obtained values reflects
the size of the cations and the stronger hydration of Mg$^{2+}$ with respect to Ca$^{2+}$.

\subsection{1:2 electrolyte solutions: lithium, sodium and potassium sulfates}
Before commenting the results for these systems, it is interesting to point out
that the properties of sodium sulfate were used to optimize the sulfate
parameters (fixing the Na parameters as obtained from NaCl systems). These
parameters were mostly fixed for the rest of sulfates. In
fact, we only modified slightly the LJ parameters of the cation-O$_s$
interaction and accepted the LB rule for other cross interactions.

\begin{center}
\begin{figure}[H]
\centering
\includegraphics*[clip,scale=0.4]{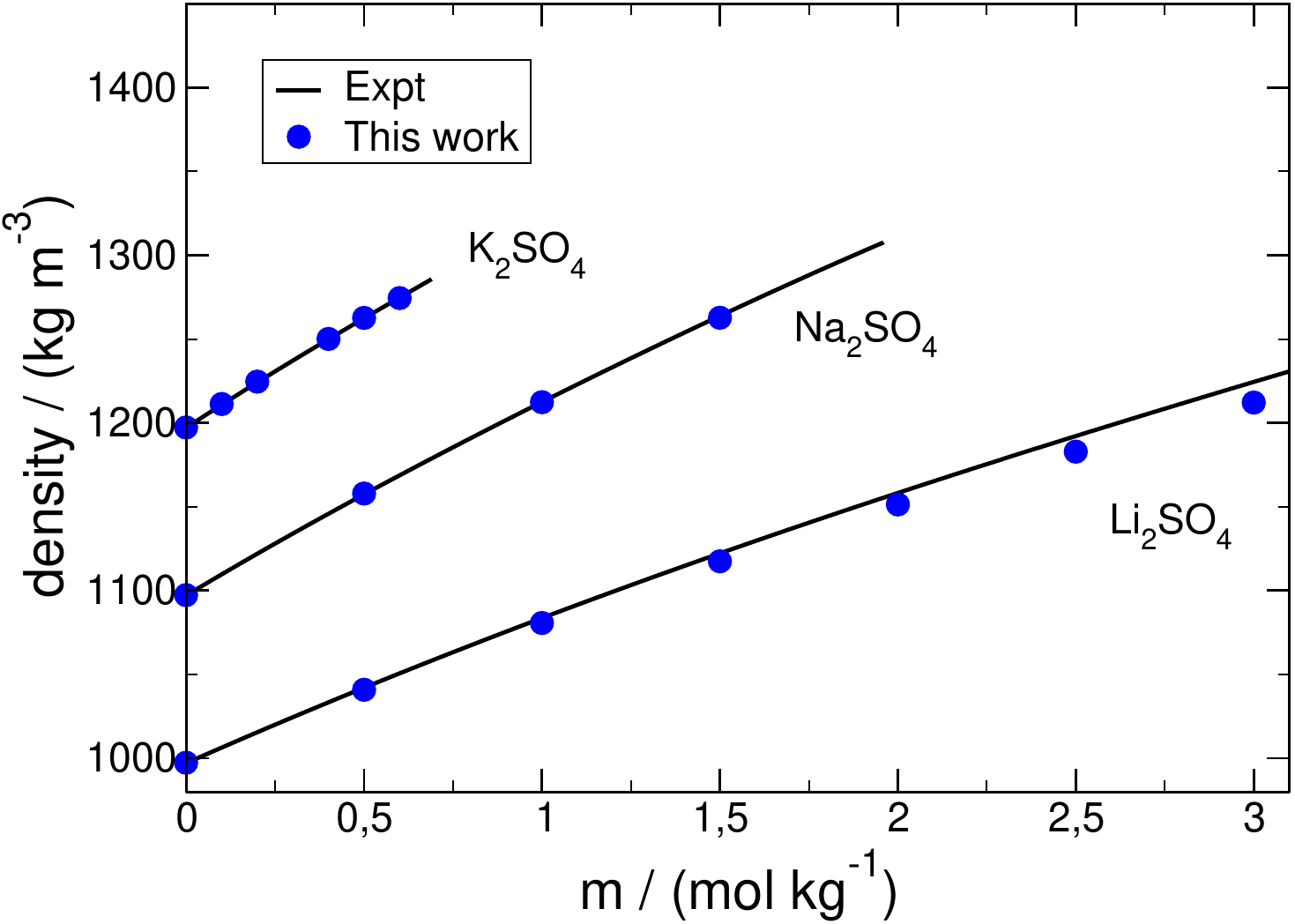}
\caption{Density as a function of salt concentration for Li$_{2}$SO$_{4}$,
Na$_{2}$SO$_{4}$, and K$_{2}$SO$_{4}$ aqueous solutions at T= 298.15 K and 0.1
MPa. Blue circles are the results from this work and the continuous line is
our fit of experimental data taken from Refs.~\onlinecite{lal:jced04,lal:jced09,mao:ap17}
and references therein.
Data for Na$_{2}$SO$_{4}$ and K$_{2}$SO$_{4}$ are shifted up 100 and 200 units,
respectively.}
\label{fig_dens_li2so4_na2so4_k2so4}
\end{figure}
\end{center}

Fig.~\ref{fig_dens_li2so4_na2so4_k2so4} shows the densities of the sulfates of
monovalent cations as predicted by the Madrid-2019 force field. They are in
excellent agreement with experiment\cite{lal:jced04,lal:jced09,mao:ap17} although small deviations are visible at
high concentrations in the case of Li$_{2}$SO$_{4}$.
Table~\ref{tab_res_sulfatos} collects the structural results obtained for 1:2
electrolytes with the sulfate as the anion.
As also observed in 1:1 electrolytes, the hydration of sulfate solutions
is strongly dependent of the cation size. No
contact ion pairs are found for the smallest cation which explains the
hygroscopic properties of lithium sulfate. The CIP of sodium sulfate is
already significant, 0.37, and that of the potassium sulfate reaches a
considerable value, 0.88. We did observe neither aggregation nor precipitation
at the highest concentration of these salts, even after long runs.
The hydration numbers and the location of the cation-O$_{w}$ peaks are almost
exactly the same as those found previously in chloride solutions.

The solvation shell of SO$^{2-}_{4}$ is formed by about 13 water molecules,
irrespective of the accompanying cation. Similar values have been reported in
previous MD simulations.\cite{cannon94,vchirawongkwin07,duv:jpcb15} In
contrast, the experimental value is much lower, about 7. Given the enormous
size of the sulfate anion, it is difficult to believe that it is hydrated by
just 7 molecules of water.\cite{duv:jpcb15} The positions of the first maximum
of the S-O$_{w}$ and the O$_{S}$-O$_{w}$ RDFs (which once again are independent
of the corresponding counterion) are in reasonable agreement with the
experimental values.
\begin{table}[H]
\caption{As in Table \ref{tab_res_1_1_elect} but for 1:2 electrolyte
solutions. Properties in solution were calculated for Li$_{2}$SO$_{4}$,
Na$_{2}$SO$_{4}$ and K$_{2}$SO$_{4}$ at 3\,m, 1.5\,m
and 0.6\,m, respectively. CIP is just the number of cations in contact with the sulfate group.}
\label{tab_res_sulfatos}
  \begin{center}
    \begin{tabular}{ c c c c c c c }
\hline
\hline
Salt             & CIP  & HN$_{c}$   & HN$_{a}$      &d$_{c-O_w}$(\AA) &d$_{S-O_w}$(\AA) & d$_{O_s-O_w}$(\AA) \\
\hline
Li$_{2}$SO$_{4}$ & 0    & 4(3.3-5.3) & 13(6.4-8.1)   & 1.84(1.90-2.25) & 3.75(3.67-3.89) & 3.00(2.84-2.95) \\
Na$_{2}$SO$_{4}$ & 0.37 & 5.5(6)     & 13(6.4-8.1)   & 2.33(2.41-2.50) & 3.75(3.67-3.89) & 3.02(2.84-2.95) \\
K$_{2}$SO$_{4}$  & 0.88 & 6.5(6-8)   & 12(6.4-8.1)   & 2.73(2.60-2.80) & 3.75(3.67-3.89) & 3.02(2.84-2.95) \\
\hline
\hline
    \end{tabular}
  \end{center}
\end{table}

A look at Table~\ref{tab_diffusion} indicate that the extrapolation
of our simulated results for the diffusion coefficient of Li$^+$, Na$^+$,
K$^+$, Mg$^{2+}$, and SO$^{2-}_{4}$  in sulfate solutions are in overall reasonable agreement with the
reported experimental data at infinite dilution (although, very likely, the
value for Li$^+$ is a bit low). 
We are not aware of experimental data on the diffusion coefficient of water in
sulfate solutions. Our results show that, in accordance with the behavior in
other electrolyte solutions, the presence of salt significantly reduces the
diffusion of the water molecules.

\begin{center}
\begin{figure}[H]
\centering
\includegraphics*[clip,scale=0.4]{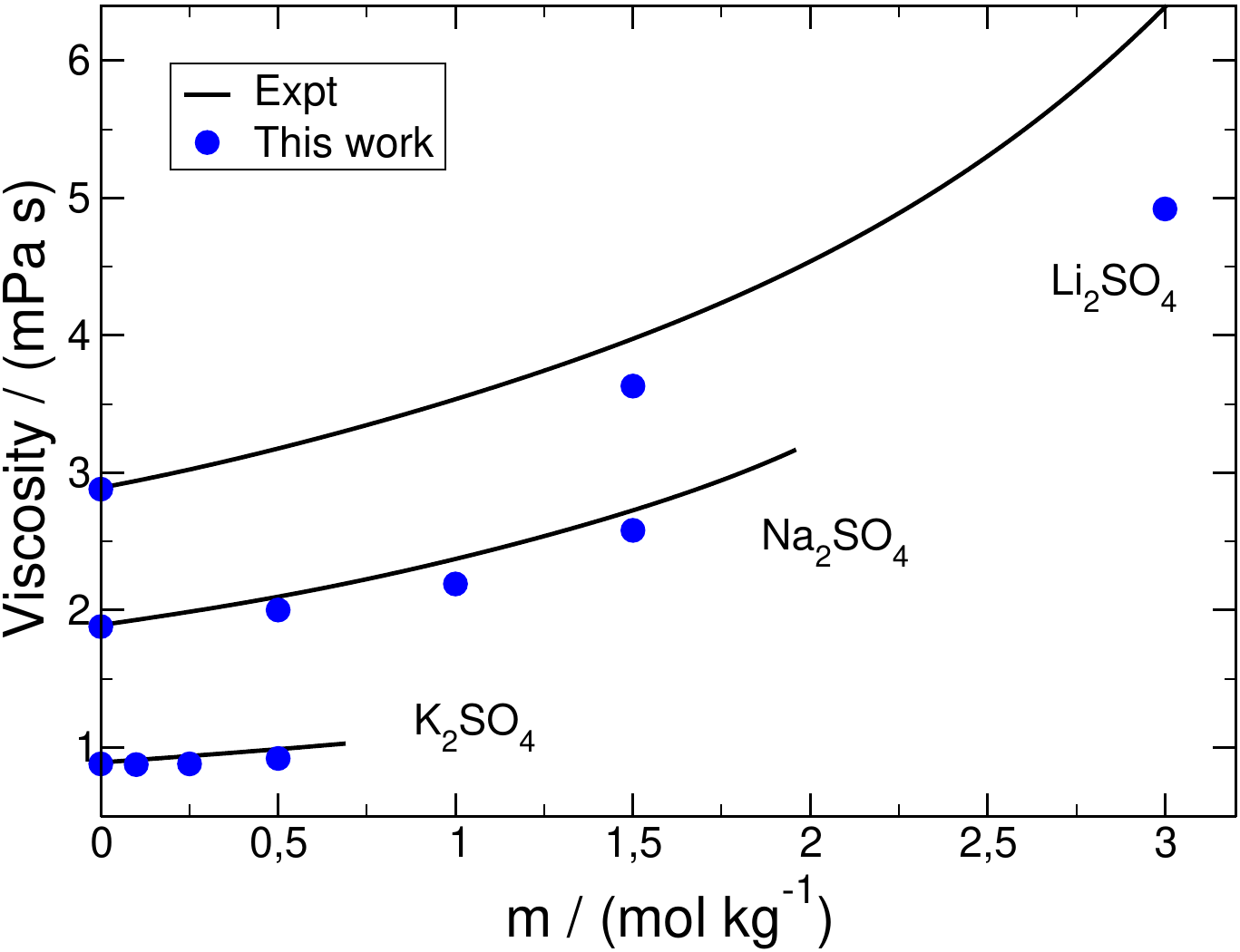}
\caption{Shear viscosity as a function of concentration for aqueous
Li$_{2}$SO$_{4}$, Na$_{2}$SO$_{4}$, and K$_{2}$SO$_{4}$ systems at 298.15 K and
0.1 MPa.  Results of this work are shown as blue circles and the continuous
line is our fit of experimental data taken from Refs.~\onlinecite{lal:jced07,lal:jced09}
and references therein.  Values for Na$_{2}$SO$_{4}$ and Li$_{2}$SO$_{4}$
solutions were shifted up one and two units, respectively.}
\label{fig_visco_li2so4_na2so4_k2so4}
\end{figure}
\end{center}

The dependence of the viscosity of sulfate solutions on the salt concentration
is shown in Fig.~\ref{fig_visco_li2so4_na2so4_k2so4}. The agreement between the
results of the Madrid-2019 force field and the experimental ones is quite good
for K$_{2}$SO$_{4}$ and Na$_{2}$SO$_{4}$. Certainly, given the relatively low
solubility of these salts, the comparison is made for concentrations where the
presence of ions does not affect dramatically the viscosity.  Our model gives
only a fair description of the viscosity of Li$_{2}$SO$_{4}$ solutions. Even at
1.5\;m concentration, the departure from experiment is much more significant
that for Na$_{2}$SO$_{4}$. Since the slope of the simulation data is largely
underestimated, the result at the solubility limit is almost one half the
experimental value.

\subsection{2:2 electrolyte solutions: magnesium sulfate}

\begin{center}
\begin{figure}[H]
\centering
\includegraphics*[clip,scale=0.4]{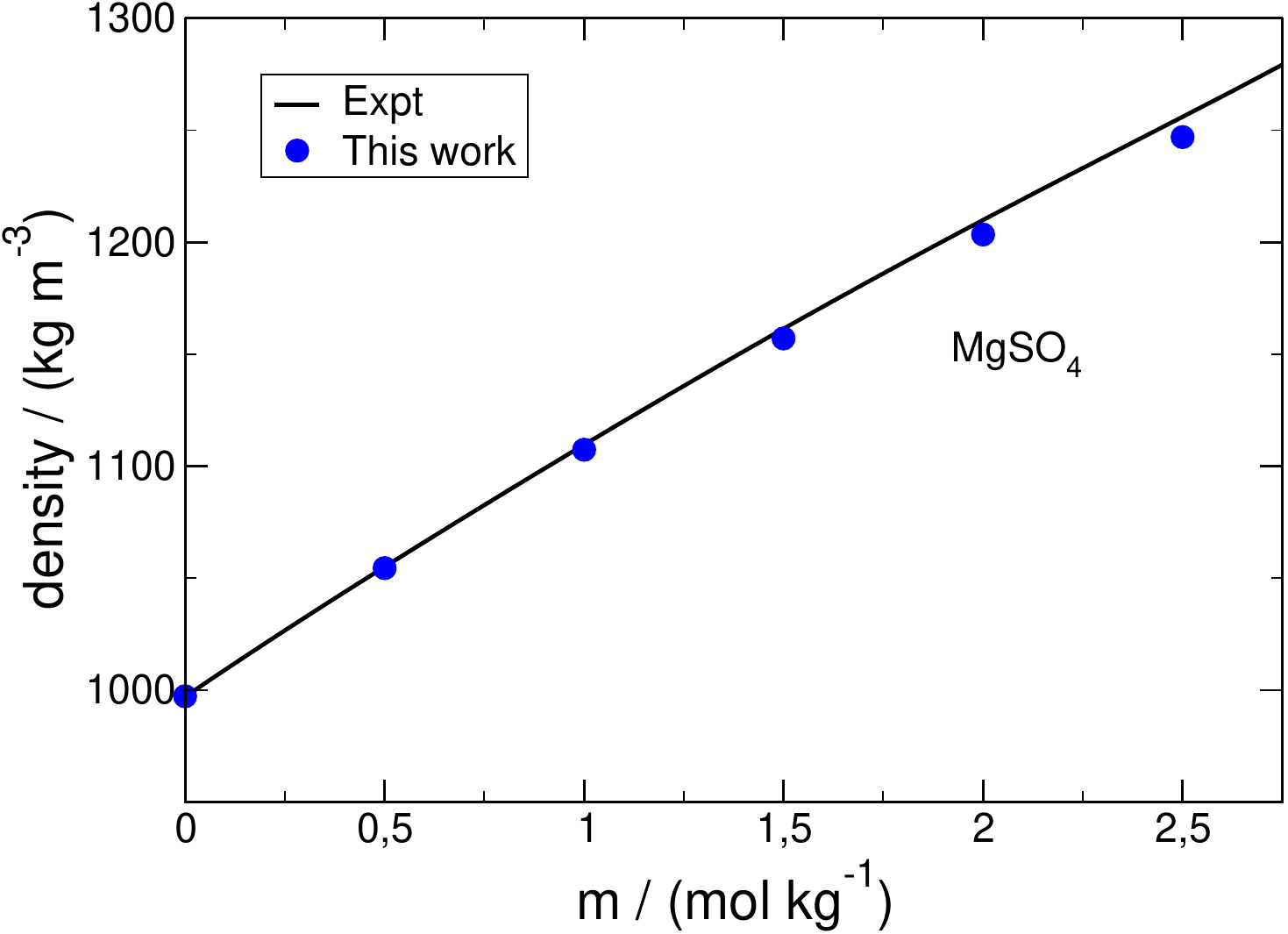}
\caption{Density as a function of molality for MgSO$_{4}$ solutions at 298.15 K
and 0.1 MPa. In blue circles, data from this force field.  The continuous line
is our fit of experimental data taken from Refs.~\onlinecite{lal:jced04,lal:jced09,mao:ap17}
and references therein.}
\label{fig_dens_mgso4}
\end{figure}
\end{center}

\begin{center}
\begin{figure}[H]
\centering
\includegraphics*[clip,scale=0.4]{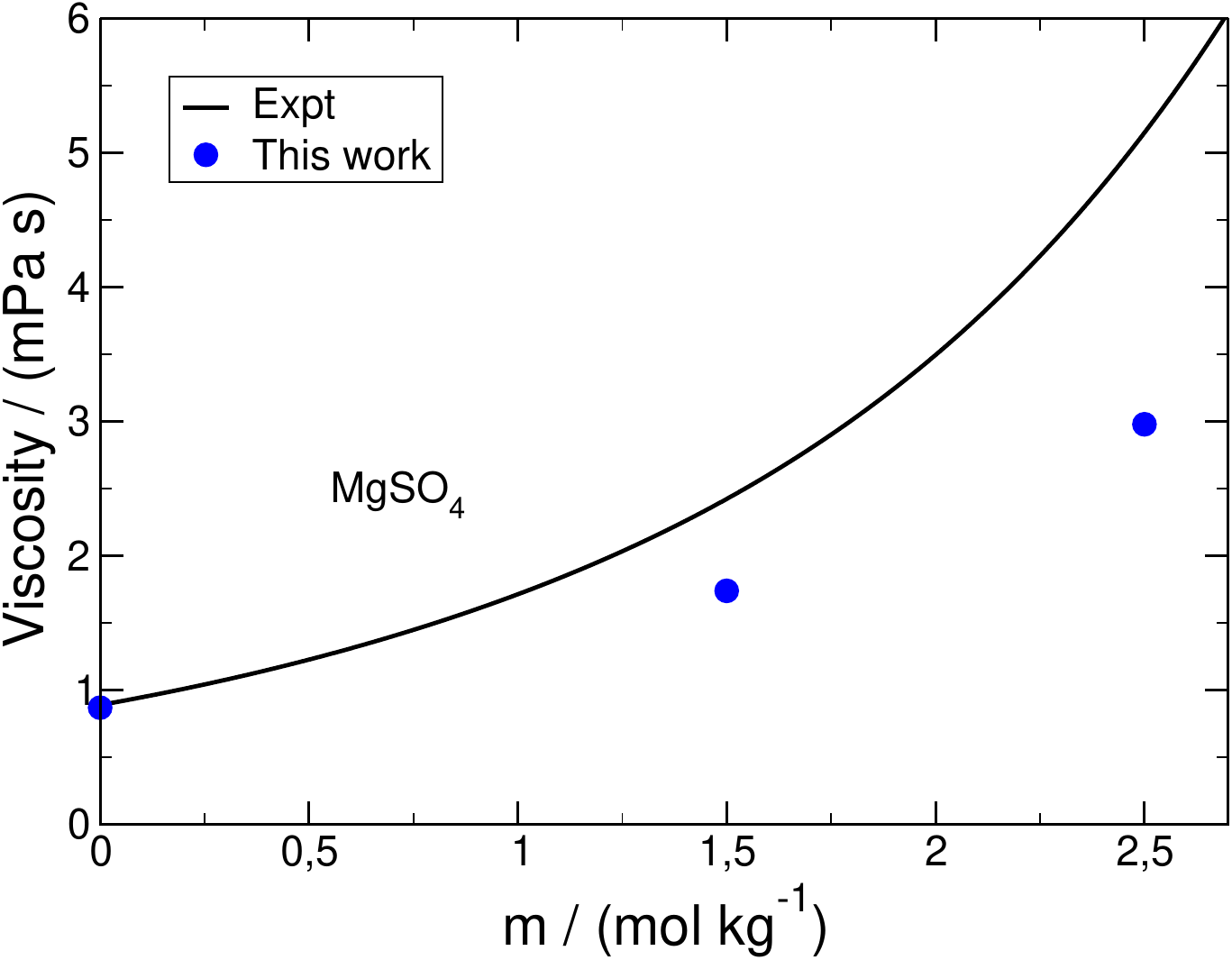}
\caption{Shear viscosity as a function of molality for MgSO$_{4}$ solutions at
298.15 K and 0.1 MPa. Blue circles are the data from this work and our fit of
experimental data taken from Refs.~\onlinecite{lal:jced07,lal:jced09} (and references
therein) is represented as a continuous line.}
\label{fig_visco_mgso4}
\end{figure}
\end{center}

The equation of state of magnesium sulfate solutions is plotted and compared to experimental results in  Fig.~\ref{fig_dens_mgso4}. The agreement is very good. Only at high salt
the simulations results underestimate the experimental density (about 1\% at 2.5\;m).  
Structural properties for a 2.5\,molal solution are presented in Table
\ref{tab_res_mgso4}. Again the magnesium is surrounded by 6 molecules of water
located at a distance of 1.92~\AA\ and the sulfate anions are surrounded by
13.5 molecules of water at 3.75~\AA. Also, the strong hydration layer
disables the formation of contact ion pairs. At this concentration no cluster
aggregation is observed in simulations of large systems for long times.
\begin{table}[H]
\caption{As in Table \ref{tab_res_1_1_elect} but for a 2.5\,m MgSO$_{4}$ solution.}
\label{tab_res_mgso4}
  \begin{center}
    \begin{tabular}{ c c c c c c c }
\hline
\hline
Salt             & CIP  & HN$_{c}$   & HN$_{a}$      &d$_{c-O_w}$(\AA) &d$_{S-O_w}$(\AA) & d$_{O_s-O_w}$(\AA) \\
\hline
MgSO$_{4}$       & 0    & 6(6)       & 13.5(6.4-8.1) & 1.92(2.00-2.11) & 3.75(3.67-3.89) & 3.00(2.84-2.95) \\
\hline
\hline
    \end{tabular}
  \end{center}
\end{table}
In Fig. \ref{fig_visco_mgso4} the viscosity for magnesium sulfate in water is
shown. The simulation results are well below the experimental ones and the
departures increase with concentration, a feature that was also found in
MgCl$_{2}$. As with other salts containing a divalent ion, the diffusion
coefficient of water in a 1\;m solution decreases significantly respect to that
of pure water (see Table~\ref{tab_diffusion}). 

 Given the low solubility of CaSO$_{4}$ (i.e 0.02m) , the densities for this salt are practically those 
of pure water and we shall not present results for this salt. 

\subsection{Ternary mixtures}

In order to test and validate the force field of this work we here present
results for ternary mixtures of electrolytes (i.e two salts and water).  We use
LB rules for those interactions that were not determined in the optimization
process (see Tables~\ref{tab_sig} and \ref{tab_eps}).  The calculations were
done for a sample made of 555 water molecules and adjusting the number of salt
molecules to the corresponding experimental concentration.  Results for the
Madrid-2019 force field are presented in Table \ref{tab_mix} together with
experimental data.\cite{zha:jced96,cam:cjc56,zha:jced97,zez:jced14,zez:jced15}
The performance of the model is excellent. Some small deviations are in
part due to the slight different concentration from the simulations as compared
to the experimental ones. Of course the difference in concentration could
be alleviated by using a larger system in the simulation.  However, since our
primary goal is to validate the proposed force field in systems not used in the
optimization process, this test is already sufficient. The main advantage of
presenting a force field is that it can deal with system of arbitrary
compositions, pressures and temperatures.
\begin{table}[H]
\caption{Densities for ternary solutions at 298.15 K and 0.1 MPa
rounded to entire units in kg/m$^{3}$. Experimental data from 
Refs.~\onlinecite{zha:jced96,cam:cjc56,zha:jced97,zez:jced14,zez:jced15}.
m$_{1}$ and m$_{2}$ are the molalities of the salts labelled as (1) and (2),
respectively. m$_{tot}$= m$_{1}$ + m$_{2}$ is the total concentration. }
\label{tab_mix}
  \begin{center}
    \begin{tabular}{c c c c c c c c c c c c c c c}
\hline
\hline
\multicolumn{3}{c}{Expt} & \, & \multicolumn{3}{c}{This work} & \, &
\multicolumn{3}{c}{Expt} & \, & \multicolumn{3}{c}{This work} \\
\cline{1-3}
\cline{5-7}
\cline{9-11}
\cline{13-15}
m$_{tot}$& m$_{1}$/m$_{2}$ &  Density & \, & m$_{tot}$& m$_{1}$/m$_{2}$ &  Density & \, &
m$_{tot}$& m$_{1}$/m$_{2}$ &  Density & \, & m$_{tot}$& m$_{1}$/m$_{2}$ &  Density \\
\hline
\multicolumn{7}{l}{NaCl (1) + LiCl (2) + H$_{2}$O} & \, &
\multicolumn{7}{l}{NaCl (1) + KCl (2) + H$_{2}$O}\\
2.106 & 0.4835 & 1057 & \, & 2.1 & 0.50 & 1053 & \, &
0.7935 & 0.3325 & 1032 & \, & 0.8 & 1/3 & 1032 \\
2.041 & 0.7253 & 1058 & \, & 2.1 & 0.75 & 1056 & \, &
2.3760 & 2.9867 & 1088 & \, & 2.4 & 3   & 1089 \\
1.976 & 1.0880 & 1059 & \, & 1.9 & 10/9 & 1054 & \, &
3.9943 & 0.3325 & 1148 & \, & 4   & 1/3 & 1149 \\
1.846 & 2.9013 & 1062 & \, & 1.9 & 2.80 & 1060 & \, &
4.6241 & 1.0010 & 1164 & \, & 4.6 & 1   & 1164 \\

\multicolumn{7}{l}{NaCl (1) + MgCl$_{2}$ (2) + H$_{2}$O} & \, & 
\multicolumn{7}{l}{NaCl (1) + CaCl$_{2}$ (2) + H$_{2}$O}\\

1.0992 & 9.9811 & 1044 & \, & 1.1 &  10     & 1043 & \, &
0.4060 & 0.3334 & 1028 & \, & 0.4 & 1/3 & 1028 \\
3.5006 & 5.9942 & 1138 & \, & 3.5 &  6      & 1136 & \, &
2.0102 & 3.0218 & 1094 & \, & 2   & 3   & 1094 \\
3.5099 & 0.1660 & 1210 & \, & 3.5 &  1/6 & 1207 & \, &
2.8055 & 1.0001 & 1158 & \, & 2.8 & 1   & 1158 \\
4.0106 & 0.3327 & 1224 & \, & 4   &  1/3 & 1221 & \, &
4.0519 & 3.0218 & 1178 & \, & 4   & 3   & 1176 \\

\multicolumn{7}{l}{NaCl (1) + Na$_{2}$SO$_{4}$ (2) + H$_{2}$O}\\

0.5997 & 0.2001 & 1062 & \, & 0.6 & 0.2 & 1062 \\
1.5000 & 1.9991 & 1092 & \, & 1.5 & 2   & 1092 \\
1.9958 & 1.0026 & 1143 & \, & 2   & 1   & 1142 \\
3.4999 & 5.9983 & 1155 & \, & 3.5 & 6   & 1154 \\
\hline
\hline
    \end{tabular}
  \end{center}
\end{table}

\section{Conclusions}
In this work a new force field (denoted as Madrid-2019) for several ions in
water has been proposed. The main conclusions are as follows :
\begin{itemize}
\item{It is possible to design a force field that, using the concept of charge
      scaling and a factor of 0.85, is able to describe with high accuracy the
      densities of a number of salt solutions.} 
\item{Viscosities are also described quite well for concentrations up to 3\,m. At
      higher concentrations deviations are clearly visible, especially for systems
      containing a high charge density (Li$^{+}$, divalent ions).}
\item{The description of the behavior of the diffusion coefficient also improves when 
      compared to systems using full ionic charges.}
\item{Complemented with the Lorentz-Berthelot rule, the force field is able to
      yield quite good predictions for ternary mixtures of electrolytes, not
      used in the optimization of potential parameters.}
\item{No precipitation or aggregation was found in the simulations performed at
      concentrations close to the experimental solubility limit. Thus the absence of 
      artefacts in the simulations is guaranteed at least up to this concentration.}  
\item{The improvement of the performance in aqueous solutions has a cost: the densities
      predicted for the crystal salts are typically a 8\% lower than the experimental ones.
      For molten salts the deviations are even larger, of the order of a 20 per cent.}
\end{itemize}

Since the use of scaled charges for ions is a relatively recent idea it is interesting 
to present some digressions on the possible limitations/possibilities  of this approach. 
 
The properties of the solid phase and/or the molten salts are defined by the
ion-ion parameters. In contrast, the electrolyte properties at lower/moderate
salt concentrations are mostly defined by ion-water parameters. Cation-anion
parameters have little contribution until one reaches high concentrations. So
interference between ion-water and cation-anion parameters should be observed
only at high concentrations. Ideally one would like to use integer charges for
the interaction between the ions when in the melt, the solid phase 
and to restrict the use of scaled charges for
the ions at low and moderate concentrations where the ions are solvated
completely and the main contribution to the properties comes from the ion-water
interactions. This is certainly possible when using polarizable force fields.
However this is not possible when using non-polarizable models. 
At this
point one should conclude that the use of the scaled charges will benefit
significantly the description of solutions at low and medium concentrations, moderately at high 
concentrations but not in the melt and/or solid phase. 

Notice also that scaled charges are used only to describe the 
potential energy surface. It is possible in principle to use different charges 
to describe the dipole moment surface both for the water molecule and for the ions. That 
was discussed in detail in Ref.\onlinecite{vegamp15}  and also anticipated in the work by Leontyev 
and Stuchebrukhov \cite{leontyev11}. It would in principle be possible to use scaled charges to describe 
the PES and integer charges to describe the DMS. This possibility will affect the way properties
that measure the response of the system to an electric field (i.e dielectric constant, electric conductivity ) 
are calculated (although we have not implemented this possibility yet  
when presenting the results of the dielectric constant in Fig.\ref{fig_dielectric_nacl}). 
Indeed, if for example ion $Cl^{-}$ passed from the anode 
to cathode then the value of the transferred charge (contributing to the flux and conductivity expression) 
should be obviously just -1, and not -0.85. All these aspects should be analyzed in more detail 
in future work.

A price to pay for the use of scaled charges is that the solvation properties (i. e., hydration free energies) 
are too low. Thus the chemical potentials of the salts in solution are smaller in absolute values when 
compared to experiment as it was discussed in our previous work for NaCl.\cite{benavides17} However it 
would possible to correct the calculated values by adding a theoretical expression denoted as the 
electronic contribution as described in detail in Ref.\onlinecite{leontyev11} (see Eq.(2.10) in this reference). 
That would bring the simulations results in closer agreement with experiment. This is the analogue of the self 
polarization energy correction proposed when the SPC/E model was introduced.\cite{spce} The water model itself did not 
reproduce the vaporization enthalpy, but it did so when the self polarization energy theoretical correction 
was included.

Overall, the first impression is that the use of scaled charges improves the description 
of ionic solutions when compared to current force fields that use full ionic charges.\cite{kan:jcp14,benavides17,mar:jcp18,panagiotopoulos_2019_mp,laage_scaling}
It would also be of interest to determine solubilities and activity coefficients of 
the current force field. The overall performance of the force field for these
properties remains to be seen.  Further work is needed to fully explore the
limits of this type of force fields.  Although we used here the scaling factor
0.85 (which yielded good results for NaCl solutions) it would be interesting to
check if other choices (i.e 0.70, 0.75, 0.80, 0.90) could led to force fields
with an overall better performance.

\section*{Supplementary Material}

See supplementary material for numerical values of densities and viscosities obtained 
in this work for several salt solutions ,for a comparison of the Lennard-Jones parameters of
the force field of NaCl of this work (Madrid-2019 model) to that proposed previously (Madrid model) 
and for a topol.top file of GROMACS including the Madrid-2019 force field. 

\section*{Acknowledgments}
 This work has been funded by 
grant FIS2016-78117-P of the MINECO and by project UCM-GR17-910570 from UCM. I.Z. thanks  CONACYT (M\'exico) for the financial support:
Convocatoria 2018 de Apoyo para Estancias Posdoctorales en el Extranjero Vinculadas a la 
Consolidaci\'on de Grupos de Investigaci\'on y Fortalecimiento del Posgrado Nacional.
We thank the anonymous reviewers of this paper for their useful comments. 

\end{document}